# Directed Diversity: Leveraging Language Embedding Distances for Collective Creativity in Crowd Ideation

Samuel Rhys Cox [†]

National University of Singapore, Singapore, samuel.cox@u.nus.edu

Yunlong Wang [†]

National University of Singapore, Singapore, yunlong.wang@nus.edu.sg

Ashraf Abdul

National University of Singapore, Singapore, ashrafabdul@u.nus.edu

Christian von der Weth

National University of Singapore, Singapore, vonderweth@nus.edu.sg

Brian Y. Lim [*]

National University of Singapore, Singapore, brianlim@comp.nus.edu.sg

## ABSTRACT

Crowdsourcing can collect many diverse ideas by prompting ideators individually, but this can generate redundant ideas. Prior methods reduce redundancy by presenting peers' ideas or peer-proposed prompts, but these require much human coordination. We introduce Directed Diversity, an automatic prompt selection approach that leverages language model embedding distances to maximize diversity. Ideators can be *directed towards* diverse prompts and *away* from prior ideas, thus improving their collective creativity. Since there are diverse metrics of diversity, we present a Diversity Prompting Evaluation Framework consolidating metrics from several research disciplines to analyze along the ideation chain — prompt selection, prompt creativity, prompt-ideation mediation, and ideation creativity. Using this framework, we evaluated Directed Diversity in a series of a simulation study and four user studies for the use case of crowdsourcing motivational messages to encourage physical activity. We show that automated diverse prompting can variously improve collective creativity across many nuanced metrics of diversity.

## CCS CONCEPTS

• Human-centered computing • Collaborative and social computing • Collaborative and social computing theory, concepts and paradigms • Computer supported cooperative work

---

[†] Co-first authors, ordered alphabetically

[*] Corresponding author

# 1 Introduction

Crowdsourcing has been used to harness the power of human creativity at scale to perform creative work such as text editing [7,21,78], iterating designs [27], , information synthesis [54], and motivational messaging [4,50,95]. In such tasks, empowering crowd workers to ideate effectively and creatively is key to achieving high-quality results. Different prompting techniques have been proposed to stimulate creativity and improve the diversity of ideas [2,27,50,95], but they suffer from ideation redundancy, where multiple users express identical or similar ideas [10,48,76,80]. Current efforts to avoid redundancy include iterative or adaptive task workflows [99], constructing a taxonomy of the idea space [40], and visualizing a concept map of peer ideas [80], but these require much manual effort and are not scalable. Instead, we propose an automatic prompt selection mechanism — **Directed Diversity** — to scale crowd ideation diversity. Directed Diversity composes prompts with one or more phrases to stimulate ideation. It helps to direct workers towards new ideas and away from existing ideas with the workflow: 1) extract phrases from text corpuses in a target domain, 2) embed phrases into a vector embedding, and 3) automatically select phrases for maximum diversity. These phrases are then shown as prompts to ideators to stimulate ideation. The phrase embedding uses the Universal Sentence Encoder (USE) [14] to position phrases within an embedding vector space . Using the embedding vectors, we calculated distances between phrases to optimally select phrases that are farthest apart from one another; this maximizes the diversity of the selected phrases. Hence, Directed Diversity guides ideators *towards* under-utilized phrases or *away* from existing or undesirable phrases.

The embedding space provides a basis to calculate quantitative, distance-based metrics to estimate diversity in selected phrases and prompts, and subsequently ideated messages. These metrics can complement empirical measurements from user studies evaluate prompts and ideations. We curate multiple measures and evaluation techniques and propose a **Diversity Prompting Evaluation Framework** to evaluate perceived and subjective creativity and objective, computed creativity, and diversity of crowd ideations. We demonstrate the framework with experiments on Directed Diversity to 1) evaluate its efficacy to select diverse prompts in a simulation study, 2) measure the perceived diversity of selected prompts and effort to generate ideas in an ideation study, and 3) evaluate the creativity and diversity of generated ideas in validation studies using quantitative and qualitative analyses. The experiments were conducted with the application use case of writing motivational messages to encourage physical activity [2,3,50,95], though we discuss how Directed Diversity can apply to other crowd ideation tasks. In summary, our contributions are:

1. We present Directed Diversity, a corpus-driven, automatic approach that leverages embedding distances based on a language model to select diverse phrases by maximizing a diversity metric. Using these constrained prompts, crowdworkers are directed to generate more diverse ideas. This results in improved collective creativity and reduced redundancy.
2. A Diversity Prompting Evaluation Framework to evaluate the efficacy of diversity prompting along an ideation chain. This draws constructs from creativity and diversity literature, metrics computed from a language model embedding, and is validated with statistical and qualitative analyses.
3. We applied the evaluation framework in a series of four experiments to evaluate Directed Diversity for prompt selection, and found that it can improve ideation diversity without compromising ideation quality, but at a cost of higher user effort.

# 2 Background and Related Work

We discuss related research on supporting crowd ideation with the cognitive basis for creative ideation, how creativity support tools help crowd ideation, and how artificial intelligence can help collective intelligence.

## 2.1 Cognitive Psychology of Creative Ideation

Different cognitive models of creativity have been proposed to explain how ideation works. *Memory-based explanation models* describe how people retrieve information relevant to a cue (prompt) from long-term memory and process it generate ideas [1,26,53,67,68]. Since retrieval is dependent on prompts, they need to be sufficiently diverse to stimulate diverse ideation [68], otherwise people may fixate on a few narrow ideas [42]. *Ideation-based models* [64] explain how individuals can generate many ideas through complex thinking processes, including analogical reasoning [36,46,61], problem constraining [84], and vertical or lateral thinking [35]. We focus on prompting to promote memory-based retrieval than these other reasoning processes. Besides cue-based retrieval and thinking strategies, other factors influence ideation creativity, such as personal traits, motivation to perform the task, and domain-relevant skills that can affect individual creativity [90]. We provide technological support to improve the creative mental process, rather than to select creative personalities, recruit domain experts, or improve task motivation. Next, we discuss how different cognitive factors have been leveraged at scale to support creative ideation with the crowd.

## 2.2 Creativity Support Tools for Crowd Ideation

Creativity Support Tools have been widely studied in HCI to enable crowdworkers to ideate more effectively and at scale [31,32]. Showing workers ideas from their peers has been very popular [18,33,79,81], but can have limited benefit to creativity if peer ideas are too distant from the ideators' own ideas [18]. Other approaches include employing contextual framing to prompt ideators to imagine playing a role for the task [69] or using avatars for virtual interactions while brainstorming [57]. While these methods focus on augmenting individual creativity, they do not coordinate the crowd, so multiple new ideations may be redundant. More recent approaches apply provide more explicit guidance to workers. IdeaHound [80] visualizes an idea map to encourage workers to focus on gaps between peer ideas, but does not inform what ideas or topics will fill the gaps. BlueSky [40] and de Vries et al. [95] use crowd or expert annotators to construct taxonomies to constrain the sub-topics for ideation, but these taxonomies require significant manual effort to construct and are difficult to scale. Chan et al. [16] employed latent Dirichlet allocation (LDA) to automatically identify topics, but this still requires much manual curation which does not scale to many topics. With Directed Diversity, we automatically extract a phrase corpus and embed the phrases as vectors, and select diverse phrases for focused prompting. We employ pre-trained language model to provide crowd ideation support, thus we next discuss how artificial intelligence can support collective intelligence.

## 2.3 Supporting Collective Intelligence with Artificial Intelligence

Collective Intelligence is defined as groups of individuals (the *collective*) working together exhibiting characteristics such as learning, judgement and problem solving (*intelligence*) [56]. Crowdsourcing is a form of collective intelligence exhibited when crowdworkers work towards a task mediated by the crowdsourcing platform. However, managing crowdwork to ensure data quality and maximize efficiency is difficult because of the nature and volume of the tasks, and varying abilities and skills of workers [97]. HCI research has contributed much towards this with interfaces to improve crowdworker efficiency, designing incentives for workers, and workflows to validate work quality [9,38,62,89,97]. Furthermore, recent developments in artificial intelligence (AI) provides opportunities to complement human intelligence to improve the quality and efficiency of crowd work [47,97], optimize task allocation [22,28], adhere to budget constraints [45], and dynamically control quality [11]. With Directed Diversity, we used AI to optimize ideation diversity by shepherding the crowd towards more desired and diverse ideation with diverse prompt selection.

## 3 Technical Approach

We aim to improve the collective diversity of crowdsourced ideas by presenting crowdworker *ideators*, with carefully selected prompts that direct them towards newer ideas and away from existing ones. The prompts presented to the *ideators* consist of one or more phrases that represent ideas that are distinct and different from prior ideas. Prompts can have one or more phrases. As a running example throughout the technical discussion and experiments, we apply our approach to the application of motivational messages for healthy physical activity, where it is important to collect diverse motivational messages [50,95]. Figure 1 shows the 3-step overall approach to extract, embed, and select phrases. We next describe each of these steps in detail.

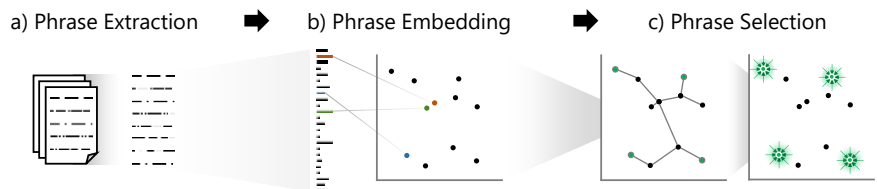

Figure 1: Pipeline of the overall technical approach to extract, embed, and select phrases to generate diverse prompts. a) Phrase extraction by collecting phrases from online articles and discussion forums (shown as pages), filtering phrases to select a clean subset (shown as the black dash for each phrase); b) Phrase embedding using the Universal Sentence Encoder [14] to compute the embedding vector of each phrase (shown as scatter plot); c) Phrase Selection by constructing the minimal spanning tree to select optimally spaced phrases (see Figure 2 for more details).

## 3.1 Phrase Extraction

We extracted phrases from selected sources of documents with the following semi-automatic data-driven process: 1) collect a corpus of documents, 2) tokenize documents into sentences, 3) extract phrases as constituent structures, 4) filter for length, slang, emoticons. We collected documents about exercising, weight loss, and healthy living from two types of sources: i) credible, authoritative health news articles[1] to obtain texts relevant to the domain (health and fitness), and ii) discussion posts from popular subreddits[2] of online health communities related to fitness and physical activity to obtain texts relevant to the task (motivational messaging [50,95]). Together, the combined corpus contained 3,235 articles and 32,721 user posts.

To extract phrases, we tokenized each document into sentences and performed Part-of-Speech (POS) tagging using Python Spacy to select phrases that form syntactic constituents [13]. From each sentence (e.g., "Regular exercising helps to improve people's health at any age.", we extract *verb phrases* (e.g., "helps to improve"), *noun phrases* (e.g., "regular exercising," "people's health," "age") and *prepositional phrases* (e.g., "at any age"). To provide more context to each phrase, we combined adjoining verb and noun phrases to generate *noun-verb phrases* (e.g. "regular exercising helps to improve") and *verb-noun phrases* (e.g., "helps to improve people's health"). After extracting the phrases, we filtered phrases for length, quality, and relevance. We kept phrases that were 3 to 5 words long, since short phrases may not sufficiently stimulate creativity and long prompts may restrict creativity. Since user posts often contain typographical errors, slang, or other stylistic devices (e.g., emoticons), we kept phrases that only contain words from a dictionary[3] of American and British words. To reduce repetition of phrases, we removed shorter phrases that overlapped with longer phrases (e.g., excluded "federal exercise recommendations", kept "federal exercise recommendations and guidelines"). The final corpus contained clean 3,666 phrases. We next describe the construction of the multi-dimensional idea space to characterize how the phrases are separated or similar.

## 3.2 Phrase Embedding

The corpus of extracted phrases provides a large set of potential phrases for prompting, but we seek to select phrases that are least similar to one another. For each phrase, we obtain a multi-dimensional vector representation, called an *embedding*, so that the phrase is a data point in an idea space. Similar work by Siangliulue et al. [79] obtained embeddings of $N = 52$ ideas by training a Crowd Kernel model [91] from 2,818 triplet annotations is not scalable to our corpus of $N = 3,666$ phrases, since that would need $N(N-1)(N-2)/3 = 16.4$ million triplets. Instead, similar to Chan et al.'s [18] use of GloVE [71], we use pre-trained language models based on deep learning to encode each word or sentence as a vector representation. Specifically, we use the more recent Universal Sentence Encoder (USE) [14] to obtain embeddings for phrases in our corpus, compute their pairwise distances, and selected a maximally diverse subset of phrases. Our approach is generalizable to other language embedding techniques [98].

Table 1: Demonstration of pairwise embedding angular distances between an example text items (first data row) and neighboring text items. Text items with semantically similar words have smaller distances. For interpretability, we highlighted words to indicate darker color with higher cosine similarity to the first phrase.

| a) Example extracted Phrases | | b) Example Ideations from Ideation User Study | |
| --- | --- | --- | --- |
| Phrase | Distance to first Phrase | Ideated Message | Distance to first Ideation |
| app with yoga poses | 0 (self) | Exercise will release endorphins and you will feel good for a while after doing it. | 0 (self) |
| yoga really taking off | 0.284 | | |
| popular form of yoga today | 0.304 | Exercise releases endorphins and makes you feel better! | 0.171 |
| yoga pants or sweats | 0.351 | Exercise relieves stress in both the mind and the body. It's the best way to get your mental health in check. | 0.301 |
| of handstand push-ups | 0.406 | | |
| on the road to diabetes | 0.475 | We are the leading country in obesity. Do you want to be part of? | 0.509 |

---

[1] Source of three authoritative websites on health: www.health.harvard.edu, www.medicinenet.com, www.webmd.com.
[2] Source of 20 subreddits from www.reddit.com: 90daysgoal, advancedfitness, advancedrunning, bodyweightfitness, c25k, crossfit ,fitness, gainit, getmotivated, ketogains, kettlebells, leangains, loseit, motivation, powerlifting, running, selfimprovement, swimming, weightroom, xxfitness.
[3] Debian Wordlist pagkage. packages.debian.org/es/sid/wordlist

To obtain the phrase embedding presentation, we use a pre-trained USE model[4] to obtain embedding vectors for each phrase. With USE, all embeddings are 512-dimensional vectors are located on the unit hypersphere, i.e., all vectors are unit length, and only their angles are different. Hence, the dissimilarity between two phrase embeddings $\boldsymbol{x}_i$ and $\boldsymbol{x}_j$ is calculated as the angular distance $\arccos(\boldsymbol{x}_i, \boldsymbol{x}_j)$, which is between 0 and $\pi$. For our phrase corpus, the pairwise distance between phrases ranged from Min=0.06 to Max=0.58, Median=0.4, inter-quartile range 0.39 to 0.46, SD=0.043; see Appendix Figure 10. We use the same USE model to compute embeddings and distances for ideated messages. For a dataset of 500 motivational messages ideated in a pilot study with no prompting, the pairwise distance between ideations ranged from Min=0.169 to Max=0.549, Median=0.405, inter-quartile range 0.376 to 0.432, SD=0.043; see Appendix Figure 11. Table 1 shows example phrases and messages and their corresponding pairwise dissimilarity distances. With the embedding vectors and pairwise distances for all phrases, the next step selects diverse phrases with which to prompt ideators.

## 3.3  Phrase Selection

Given the embeddings of the curated phrases, we want to select the subset of phrases with maximum diversity. Mathematically, this is the dispersion problem or diversity maximization problem of "arranging a set of points as far away from one another as possible". Among several diversity formulations [20], we choose the Remote-MST diversity formulation [37] (also called Remote-tree [20] or functional diversity [72]) that defines diversity as the sum of edge weights of a minimum spanning tree (MST) over a set of vertices. It is robust against nonuniformly distributed data points (e.g., with multiple clusters, see Table 4). We construct the minimum spanning tree by performing agglomerative hierarchical clustering on the data points with single linkage [82]. Next, we describe how we select phrases as prompts to direct ideators towards diverse phrases, or away from prior ideas. Figure 2 illustrates the technical approach.

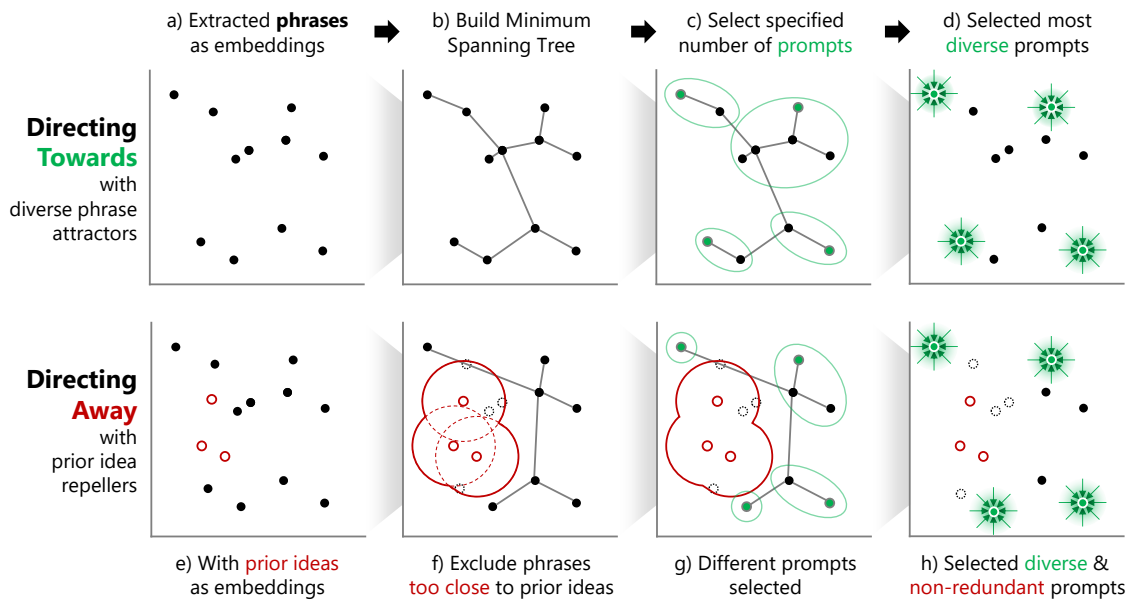

Figure 2: Procedure to direct ideation towards diverse phrases (top) and away from prior or redundant ideas (bottom). To attract ideation with diverse prompts: a) start with embeddings of corpus-extracted phrases; b) construct minimum spanning tree (MST); c) traverse tree to select distant prompts from clusters (most distant points as green dots, in clustered phrases as green ellipses); d) selected prompts are the most diverse. To repel ideation from prior ideas, e) compute embeddings of prior ideas (red hollow dots); f) compute prompt-ideation pairwise distances of all prompts from each prior ideation, exclude phrases (dotted black circles) with pairwise distance less than a user-defined threshold (red bubble), and construct the MST with remaining phrases; g) traverse MST to select a user-defined number of prompts; h) selected prompts are diverse, yet avoids prior ideas.

---

[4] Pre-trained Universal Sentence Encoder model (https://tfhub.dev/google/universal-sentence-encoder/4) , which was trained using both unsupervised learning on Wikipedia, web news, web question-answer pages, and discussion forums, and supervised learning on Stanford Natural Language Inference (SNLI) corpus.

### 3.3.1 Directing towards Diverse Phrases

For phrase selection, we aim to select a *fixed number of points* $n$ from the corpus with maximum diversity. This is equivalent to finding a maximal edge-weighted clique in a fully connected weighted graph, which is known to be NP-hard [39]. Hence, we propose a scalable greedy approach that uses the dendrogram representation of the MST resulting from the hierarchical clustering. Starting from the root, we set the number of clusters to the desired number of phrases $n$. For each cluster $C_r$, we select the phrase that is most distant from other points, with largest minimum pairwise distance from all points from outside the cluster, i.e.,

$$x_r = \underset{i \in C_r}{\mathrm{argmax}} \left( \min_{j \notin C_r} d(x_i, x_j) \right)$$

where $x_r$ is the diverse phrase selected in cluster $C_r$, $x_i$ is a point in cluster $C_r$ and $x_j$ is a point in the corpus not in $C_r$, and $d$ is the pairwise distance between $x_i$ and $x_j$. This method has $O(n^2)$ time complexity and runs in less than one second on a desktop PC for 3.6k phrases; it is generalizable and can be substituted for other approximate algorithms to select most diverse points [20,41]. Figure 2 (top row) illustrates the phrase selection method to direct towards areas without ideations:

    a) Start with all phrases in a corpus represented as USE embedding points.
    b) Construct a dendrogram (MST) from all points, using single-linkage hierarchical clustering.
    c) Set # clusters equal to desired number of diverse phrases. For each cluster, find the most distant phrase.
    d) Selected phrases are the approximately most diverse from the corpus, for the desired number of phrases.

### 3.3.2 Directing Away from Prior Ideas

Other than directing ideators towards new ideas with diverse prompts, it is important to help them to avoid prior ideas written by peers. We further propose a method to remove corpus phrases that are close to prior ideas so that ideators do not get prompted to write ideas similar to prior ones. The method, illustrated in Figure 2 (bottom row), is similar as before, but with some changes:

    e) Add the embedding points of prior ideas to the corpus.
    f) Calculate phrase-ideation distance $d(x_i^P, x_j^I)$ for each phrase $x_i^P$ and ideation $x_i^I$ and exclude phrases too close to the ideas, i.e., $d < \delta$, where $\delta$ is an application-dependent threshold, $\delta = 0.29$ in our case.
    g) Same as step (c), but different clusters, since fewer points are clustered.
    h) Same as step (d), but different prompts would be selected, even if the number of phrases is the same.

### 3.3.3 Directing with Prompts of Grouped Phrases

Instead of prompting with only one phrase, prompting with multiple related terms can help ideators to better understand the concept being prompted and generate higher quality ideas [17,67,83]. We extend the phrase selection method to group multiple phrases in a single prompt using the following greedy algorithm. After step (a), we i) sorted phrases by descending order of minimum pairwise distance for each phrase to produce a list of seed candidates, ii) for each seed phrase, perform a nearest neighbors search to retrieve a specified prompt size (number of phrases $g$ in a prompt) and remove the selected neighbors from the seed list, iii) repeat seed neighbor selection until $n$ seed phrases have been processed. We grouped the phrases into a prompt and calculate its embedding point $x_i^{Pr}$ as the angular average of all phrases $x_k^P$ in the prompt, i.e., $x_i^{Pr} = \sum_{k=1}^{g} x_k^P / Z$, where $Z = \left\| \sum_{k=1}^{g} x_k^P \right\|_2$ is the magnitude of the vector sum and $x_i^{Pr}$ is also a unit vector. We then perform steps (b) to (d) with the prompts $x_i^{Pr}$ instead of individual phrases. Note that the corpus of prompts will be smaller than the corpus of phrases. This approach has disjoint prompts that do not share phrases, but there can be alternative approaches to group phrases[5].

## 4 Diversity Prompting Evaluation Framework

To evaluate the effectiveness of the Directed Diversity prompt selection technique to improve the collective creativity of generated ideas, we define an ideation chain as a four step process (Figure 3 top): 1) setting the prompt selection technique will influence 2) the creativity of selected prompts (prompt creativity), 3) the ideation process of the ideators (prompt-ideation mediation), and 4) the creativity of their ideation (ideation creativity). We propose a Diversity Prompting Evaluation Framework, shown in Figure 3, to measure and track how creative and diverse information propagates along this ideation chain to evaluate how and whether a creativity prompting technique improves various measures of creativity and diversity in outcome ideas. Note that our proposed framework is descriptive to curate many useful metrics, but not prescriptive to recommend best metrics.

---

[5] An alternative approach is, after step (c), to simply group nearest neighbors. However, this will cause the prompt embeddings to be shifted after the diversity is maximized, so it may reduce the diversity of the selected prompts.

## 4.1 Research Questions and Experiments

Prompt stimuli act along the ideation chain to increase ideation diversity, but it is unclear how well they work and at which point along the chain they may fail. We raise three research questions between each step in the ideation chain, which we answer in four experiments (Section 5) with various measures and factors.

**RQ1.** *How do the prompt techniques influence the perceived diversity of prompts? (RQ1.1) How do they affect diversity in prompts? (RQ1.2) How well can users perceive differences in creativity and diversity in these prompts?* These questions relate to the prompt selection technique effectiveness and serve as a manipulation check. We answer them in a Characterization Simulation Study (Section 5.1) with objective diversity measures, and an Ideation User Study (5.2) with subjective measures perceived prompt diversity measures.

**RQ2.** *How does diversity in prompts affect the ideation process for ideators? (RQ2.1) Do differences in diversity affect ideation effort? (RQ2.2) How well do ideators adopt and apply the content of the prompts? (RQ2.3) How does prompt creativity affect diversity in ideations?* We answer these questions as a mediation analysis in the Ideation User Study (Section 5.2) with objective measures of task time and similarity between ideations and stimulus prompts, thematically coded creativity metrics, and perceived ease of ideation.

**RQ3.** *How do prompt selection techniques affect diversity in ideations?* Having validated the manipulation checks, we evaluate the effectiveness of prompt selection techniques in questions in the Ideation User Study (Section 5.2) with subjective measures self-assessed creativity and thematically coded creativity metrics, and two Validation User Studies (Section 5.3) with subjective measures of perceived creativity.

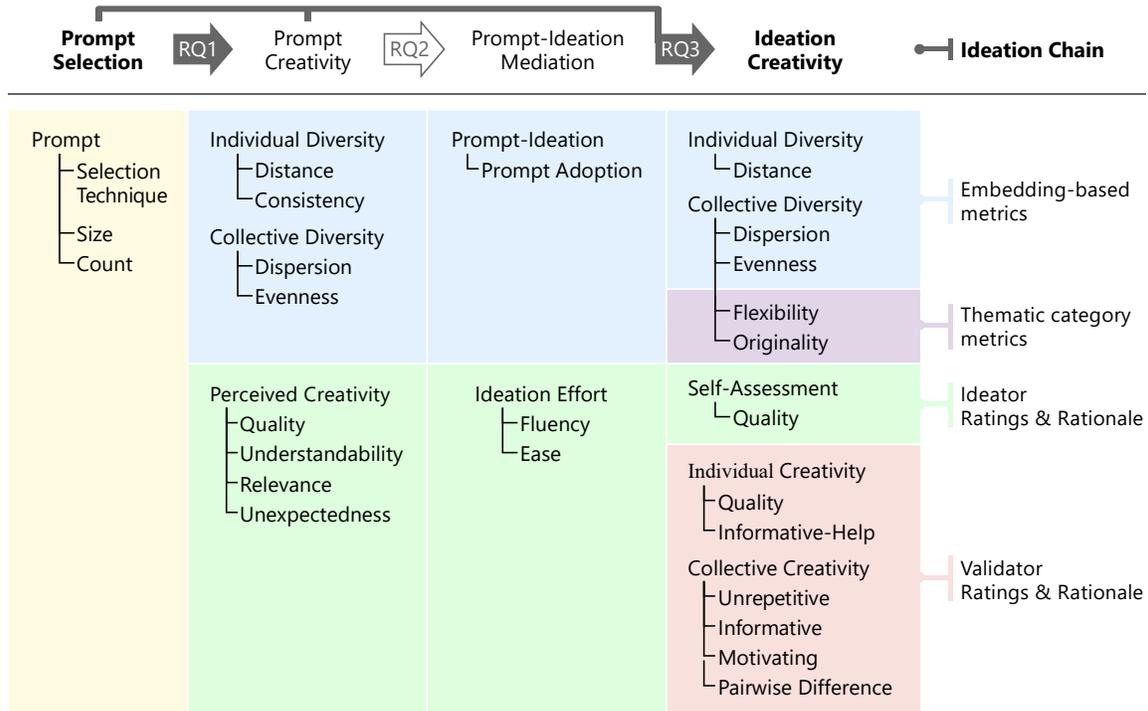

Figure 3: Diversity prompting evaluation framework to evaluate prompting to support diverse ideation along the ideation chain. We pose research questions (RQ1-3) between each step to validate the ideation diversification process. For each step, we manipulate or measure various experiment constructs to track how well ideators are prompted to generate creative ideas. Except for prompt selection, each construct refers to a statistical factor determined from factor analyses of multiple dependent variables. Constructs are grouped in colored blocks indicating different data collection method (☐ Computed embedding-based metric, ☐ ratings from ideators, ☐ ratings from validators, ☐ thematic coding of ideations).

### 4.1.1 Independent variables of Prompt Specifications

We manipulated *prompt selection* technique, *prompt count*, and *prompt size* as independent variables; these are detailed in Appendix Table 6. We chose Random prompt selection as a key baseline where selection is non-trivial and data-driven based on our corpus, but not intelligently selected for diversity.

## 4.2 Diversity and Creativity Measures of Prompting and Ideation

We measured diversity and creativity for selected prompts and generated ideas with embedding-based and human rated metrics. We color code variable names based on data collection method as in Figure 3.

### 4.2.1 Embedding-based Diversity Metrics for Prompts and Ideations

Although crowd creativity research has focused on the mean pairwise distance as a metric for idea diversity, our literature review has revealed many definitions and metrics. Here, we describe computational metrics calculated from the embedding-based distances. Inspired by Stirling's general framework diversity framework [87], we collect definitions from crowd ideation [15,27,40,79,80], ecology [24,73,94], recommender systems [29,44,60,93], and theoretical computer science [20,37]. These cover many aspects of diversity to characterize the *mean distance* and minimum *Chamfer distance* between points, *MST-based dispersion*, *sparseness* of points around the median, *span* from the centroid, and *entropy* to indicate the evenness of points in the embedding vector space. Table 2 and Table 3 describe distance metrics for individual and collective text items, respectively. These metrics describe nuances of diversity, which we illustrate with example distributions in Table 4. Other measures of diversity and divergence [20] can be included in the framework, which we defer to future work. Next, we describe human-subjects ratings to validate these embedding-based metrics with measures that do not depend on the embeddings to avoid circular dependency.

Table 2: Metrics of distances between two points in a multi-dimensional vector space. Each metric can be calculated for an individual text item. These metrics can apply to the embedding of phrases or ideations.

| Metric | Definition | Interpretation |
| --- | --- | --- |
| *Mean Pairwise Distance* | $\frac{1}{N-1}\sum_{j=1}^{N} d(\pmb{x}_i, \pmb{x}_j)$ | Average distance of all other points to the current point. |
| *Minimum Pairwise Distance* | $\min_{j \neq i} d(\pmb{x}_i, \pmb{x}_j)$ | Distance of closest neighbor to current point. This focuses on redundancy and ignores points that are very far from the current point. |

Table 3: Metrics of diversity of phrases or ideation embeddings in a vector space. These capture more characteristics of diversity than average distances in Table 2. Each metric can only be calculated collectively for multiple items.

| Metric | Definition | Interpretation |
| --- | --- | --- |
| *Remote-Clique* | $\frac{1}{N^2}\sum_{i,j} d(\pmb{x}_i, \pmb{x}_j)$ | Average of mean pairwise distances. While commonly used in crowd ideation studies [27,44,80], it is insensitive to highly clustered points. |
| *Chamfer Distance* | $\frac{1}{N}\sum_{i=1}^{N} \min_{j \neq i} d(\pmb{x}_i, \pmb{x}_j)$ | Average of minimum pairwise distances. Chamfer distance [43] (or Remote-pseudoforest [20]) measures the distance to the nearest neighbor. However, it is biased when points are clustered. |
| *MST Dispersion* | Mean of MST edge distances $\frac{1}{|E_{MST}|}\sum_{(x_i,x_j)\in E_{MST}} d(\pmb{x}_i, \pmb{x}_j)$ | Popular in ecology research as functional diversity [72], and called Remote-tree or Remote-MST [20,37], this learns a minimum spanning tree (MST) of the points, and calculates the sum of edge weights. |
| *Span* | $\text{percentile}_{P\%}\, d(x_i, \bar{x}^M)$ | $P^{\text{th}}$ percentile distance to centroid ($\bar{x}^M = \sum_{i=1}^N x_i^M / N$); i.e., "radius" of distribution [12,65]. We calculate 90th percentile to centroid (vs. medoid) to be robust against outliers and skewed distributions, respectively. |
| *Sparseness* | Mean distance to medoid $\frac{1}{N}\sum_{i=1}^{N} d(x_i^M, \tilde{x}^M)$ | Sparsity of points positioned around the medoid ($\tilde{x}^M = \text{argmin}_{x_i}\{\sum_{j=1}^{N} d(\pmb{x}_i, \pmb{x}_j)\}$) [51,52,77]. If points cluster around the medoid, then this metric will be small (i.e., not sparse). |
| *Entropy* | Shannon-Wiener index for points in a grid partition $\sum_b f_b \log(f_b)$ | This index [75,86] indicates how evenly points are distributed; more even is more diverse. We calculated entropy for a 2D projection of the USE feature space to avoid high time complexity[6] and divided the space into a 5×5 grid[7], and counted the frequency $f_b$ of points in each bin $b$. |

---

[6] Since calculating entropy in high dimensions is computationally expensive, we reduce the 512-dimension USE feature space to a 2-dimension UMAP projection [59]. This is a dimensionality reduction technique that is more robust than t-SNE. We iterated hyperparameters settings and chose the projection with highest correlation between the entropy results and mean pairwise distances.

[7] Entropy calculations will differ for different grid sizes, but the general trends with respect to points distribution should be similar.

Table 4: Comparison of diversity metrics for canonical examples of distributed points in a 2D space. Points farther apart mean higher diversity. Here, we calculate Euclidean instead of angular distance, but intuitions are similar.

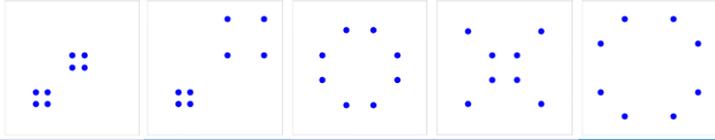

| | | | | | |
|---|---|---|---|---|---|
| Remote-Clique | 0.258 | 0.448 | 0.411 | 0.389 | 0.561 |
| Chamfer Distance | 0.800 | 1.600 | 1.680 | 1.931 | 2.263 |
| MST Dispersion | 0.126 | 0.232 | 0.244 | 0.247 | 0.333 |
| Span | 0.218 | 0.370 | 0.327 | 0.283 | 0.447 |
| Sparseness | 0.221 | 0.369 | 0.409 | 0.303 | 0.561 |
| Entropy | 0.693 | 1.386 | 1.733 | 1.386 | 2.079 |

#### 4.2.2 Creativity Measures for Ideations

Along with the computed diversity metrics, we evaluate with qualitative characteristics of creativity. From creativity literature, we draw from Torrance's [92] description of several measures for creativity, including quality, flexibility and originality. *Quality* measures whether an ideation is "usable, practical, or appropriate" [66]. We asked ideators to self-assess on a 5-point Likert scale their message's *effectiveness* (towards motivation) and *creativity*. We ask validator crowdworkers to rate each individual ideation on a 7-point Likert scale whether it is *effective* (motivating [95]), *helpful*[8] [50,88], and *informative* [50] towards encouraging physical activity; rank collections of ideations on *effectiveness*, *informativeness* and *unrepetitiveness*.; and rate the *pairwise difference* between ideation pairs from each collection. Note that Directed Diversity was not designed to improve quality, since these metrics were not explicitly modeled. *Flexibility* [85] measures how many unique ideas were generated, and *originality* [100] measures how infrequently each idea occurs. These require expert annotation to identify distinct categories. We conducted a thematic analysis on the messages using open coding [34] to derive categories and affinity diagramming [8] to consolidate categories to themes (see details in Appendix Table 19). We calculate the flexibility and originality measures based on the coded categories (fine-grained) and themes (coarser) described in Appendix Table 7.

#### 4.2.3 Creativity Measures for Prompts

As a manipulation check, it is important to verify that prompts that are computed as more diverse, are perceived by ideators as more creative. Since perceived creativity encompasses more qualitative effects, computed diversity may not be correlated with creativity. Thus, we measure the creativity and usefulness of prompts by asking about prompt *understandability, relevance to domain topic* (physical activity), *relevance to task*[9] (motivation), *helpfulness* to inspire ideation, and *unexpectedness* [66] along 7-point Likert scales.

#### 4.2.4 Mediating Variables for Prompt-Ideation Process

Even if more diverse prompts can facilitate more creative ideation, it is important to understand whether this requires more effort and time, how the consistency of phrases within prompts affect ideation, and how well ideators adopt words and concepts from the phrases into their ideations. We measure effort as *ease of ideation* with a 7-point Likert scale survey question. For individual creativity, fluency [30] is defined as the number of ideas an individual writes within a fixed time. Chan et al. had also measured fluency for an 8-minute crowd ideation task [18]. In contrast, we asked ideators to only write one idea per prompt without time constraint, so we measure the inverse relation of ideation task time to generate one ideation [5]. Specifically, since task time is skewed, we use $-Log(ideation\ time)$ to represent *fluency*. For prompts with more than one phrase, the similarity between phrases can affect their perceived consistency. Therefore, we measure the *intra-prompt mean phrase* and *prompt average phrase Chamfer distances* (Appendix Table 8) to indicate the similarity between intra-prompt phrases. We measure the adoption of the prompt ideas by calculating the proportion of words from phrases in the ideations as *prompt recall* and *prompt precision*, and computing the *prompt-ideation distance* between the embeddings of the prompt and ideation (Appendix Table 9).

---

[8] Note that a message could be helpful but written with negative impressions and thus not motivating.
[9] Note that a prompt could be relevant to the domain, but not motivating.

## 4.3 Factor analyses to draw constructs from experiment variables

With the numerous variables from our experiments, we observed some may be correlated since they measure similar notions or participants may confound questions to have similar meanings. We employed an iterative design-analytical process to organize and consolidate variables into factors with the following steps.

- Identify metrics of creativity and diversity from a literature review from various research domains, such as ecology, creativity, crowdsourcing, theoretical computer science, recommender systems (Section 4.2.1). Ideate additional measures and questions to capture user behavior and opinions when generating and validating ideas. We refine and reduce measures based on survey pilots and usability testing.
- Collect measurements of each metric with different methods: a) Compute **embedding-based metrics** from prompts shown and messages written. This was computed *individually* for each text item (e.g., mean pairwise distance) and *collectively* for all text items in each prompt technique (e.g., Remote-MST diversity). b) Measure **perception ratings and behavioral measures** regarding reading prompts and ideating messages and rating messages. We asked text rationale to help with interpretations. c) Measure **subjective thematic measures** to qualitatively assess the collective creativity with thematic analysis and idea counting.
- Perform factor analysis on quantitative data to organize correlated variables into fewer factors. Variables are first grouped by data collection method[10] and analyzed together. To determine the number of factors, we examined scree plots and verified grouped variables as consistent with constructs from literature. The final number of factors are statistically significant by the Bartlett Test of Sphericity (all $p<.0001$). See Appendix Tables 10-17 for the results of the factor analysis, including factor loadings and statistical significance. Table 5 summarizes the learned factors from 42 variables that we developed.
- Perform statistical hypothesis testing using these learned factors to answer our research questions.

Table 5: Constructs from factor analyses of variables along ideation chain. Factor loadings in Appendix Tables 10-17.

| Chain | Factor Construct | Interpretation |
|---|---|---|
| Prompt Creativity | Prompt Distance | How distant and isolated the prompt is from other prompts. |
| | Prompt Consistency | How similar (consistent) the phrases in a prompt are. |
| | Prompt Dispersion | How spread out the selected prompts are from one another. |
| | Prompt Evenness | How evenly spaced the selected prompts are among themselves. |
| | Prompt Unexpectedness | Ideator rating of how unexpected a prompt was on a 5-pt Likert scale. |
| | Prompt Understandability | Ideator rating of how understandable a prompt was on 5-pt Likert scale |
| | Prompt Relevance | Ideator rating of prompt relevant to the domain (i.e. exercise) on 5-pt Likert scale |
| | Prompt Quality | Ideator rating of the overall quality of prompt on 5-pt Likert scale. |
| Prompt-Ideation Mediation | Ideation Fluency | Ideator speed to ideate (reverse of time taken). |
| | Ideation Ease | Ideator ease of ideating based on multiple 5-point Likert scale ratings. |
| | Phrase Adoption | Measures the extent of phrase usage from the prompts in the ideation. |
| Ideation Creativity | Ideation Distance | How distant and isolated the ideation is from other ideations. |
| | Ideation Dispersion | How spread out the ideations are from one another. |
| | Ideation Evenness | How evenly spaced the ideations are among themselves. |
| | *Ideation Flexibility* | Count of unique categories/themes across all ideations. |
| | *Ideation Originality* | How rare each category/theme is across all ideations. |
| | Ideation Self-Quality | Ideator self-rating of the overall quality of the ideation on 5-pt Likert scale. |
| | Ideation Quality | Validator rating of overall quality of individual ideation on 7-pt Likert scale. |
| | Ideation Informative-Helpfulness | Validator rating of informativeness and helpfulness of individual ideation on 7-pt Likert scale. |
| | Ideations Unrepetitive | Validator cumulative rating of non-redundancy in collection of ideations. |
| | Ideations Informative | Validator cumulative rating of informativeness in collection of ideations. |
| | Ideations Motivating | Validator cumulative rating of overall quality of collection of ideations. |
| | Ideations Pairwise Difference | Validator rating of difference between a pair of ideations in collection. |

---

[10] E.g., individual text item metrics, collective text items metrics, ratings of text item from ideators, ratings of text item from validators, ratings of collection of text items from validators.

# 5 Evaluation: applying framework to study Directed Diversity

We have described a general descriptive framework for evaluating diversity prompting. We applied it to evaluate our proposed Directed Diversity prompt selection technique against baseline approaches (no prompting, random prompt selection) in a series of experiments (characterization, ideation, individual validation, collective validation), for the use case of crowd ideating motivational messages for physical activity. Here, we describe the procedures for each experiment and their results.

## 5.1 Characterization Simulation Study

The first study uses computational methods to rapidly and scalably evaluate prompt selection techniques. This helps us to fine tune prompt parameters to maximize their potential impact in later human experiments.

### 5.1.1 Experiment Treatments and Method

We varied three independent variables (prompt selection, prompt count, prompt size) to measure the impact on 7 dependent variables of distance and diversity metrics. We varied *Prompt Selection* technique (None, Random, or Directed) to investigate how much Directed Diversity improves prompt diversity with respect to baseline techniques. For None prompt selection, we simulated ideation with 500 ideas collected from a pilot study where crowd ideators wrote messages without prompts. We simulated Random selection by randomly selecting phrases from the phrase corpus (Section 3.1) and Directed selection with our technical approach (Sections 3.1 to 3.3). If we assume that prompt embeddings are an unbiased estimator for ideation embeddings, then this gives an approximation of ideation diversity due to prompting. We conducted experiments for directing towards diverse prompts and for directing away from the 500 pilot prior ideations. We varied the number of prompts (*Prompt Count*, $n = 50, 150, \dots, 950$) to simulate how diversity increases with the number of ideation tasks performed. This investigates how diversity increases as the budget for crowd tasks increases. To investigate how well Directed selection avoids prior ideations, we varied the number of repeller prior ideations (Repeller Prior Ideations Count, $n_R = 50, 100, 150, 200$). We varied the number of phrases in prompts (*Prompt Size*, $g = 1$ to 5) to simulate ideating on one or more phrases in each prompt. We computed the prompt embedding as the average of all phrases in the prompt. For Random selection, we randomly chose phrases to group together for each prompt. This random neighbor selection will lead to variation in prompt consistency, but does not bias the prompt embedding on average. For Directed selection, phrases in each prompt were chosen as described in Section 3.3.3.

### 5.1.2 Results on Manipulation Efficacy Analysis (RQ1.1)

We visualized (Figure 4) the phrase embeddings to help to interpret how the selected prompts are distributed, whether they are well spread out, clustered, etc. We used Uniform Manifold Approximation and Projection (UMAP) [59] to reduce the 512 dimensions of USE to a 2D projection. Hyperparameters were selected such that the 2D points in UMAP had pairwise distances correlated with that of the 512-dimension USE embeddings.

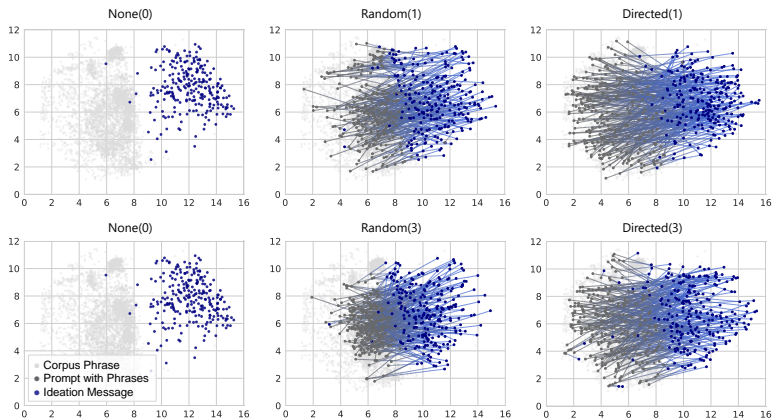

Figure 4: 2D UMAP projection showing how diversely selected prompts and resulting ideation messages are distributed based on Directed or Random prompt selection technique and prompt size (number in brackets). Each point represents the embedding of a text item. Light grey points represent all phrases in the extracted corpus, dark grey points represent selected phrases from the simulation study (Section 5.1) and blue dots represent the ideated messages written by crowdworkers in the ideator user study (Section 5.2). Gradient lines connect ideation messages to their stimulus prompts.

We can see that Directed prompt selection led to prompts that were more spread out, and less redundant from prior ideation. This is more pronounced for higher prompt size ($g = 3$). Random(3) had lower diversity than None with tighter clustering of prompts (grey points in middle-bottom graph) than of messages (blue points in left graph). This was because Random(3) prompts averaged their embeddings from multiple phrases, such that this variance of means of points is smaller than the variance of points[11]. We further conducted a characterization study with 50 simulations for each prompt configuration to confirm that Directed Diversity improves diversity and reduces redundancy from prior ideations for various embedding-based metrics (see Appendix E and Figure 12).

## 5.2 Ideation User Study

The Ideation User Study serves as a manipulation check that higher prompt diversity can be perceived by ideators, and as an initial evaluation of ideation diversity based on computed and thematically coded metrics.

### 5.2.1 Experiment Treatment and Procedure

We conducted a between-subjects experiment with two independent variables prompt selection technique (None, Random, Directed) and prompt size ($g = 1$ and 3), and kept constant prompt count $n = 250$. The None condition (no prompt) allows us to measure if the quality of ideations become worse due to the undue influence of phrases in prompts. The Random condition provides a strong baseline since it also leverages the extracted phrases in the first step of Directed Diversity. A prompt size of $g > 1$ can provide more contexts to help ideators understand the ideas in the phrases, but may also lead to more confusion if the phrases are not consistent (too dissimilar). Figure 5 shows example prompts that ideator participants see in different conditions. The experiment apparatus and survey questions were implemented in Qualtrics (see Appendix Figures 13-19 for instructions and question interface).

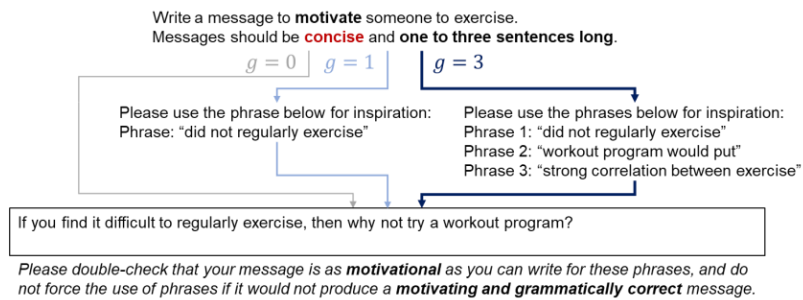

Figure 5: Example prompts shown to participants in different conditions: None (left, $g = 0$), Directed(1) (center, $g = 1$), and Directed(3) (right, $g = 3$). Phrase texts would be different for Random(1) and Random(3) selection techniques.

### 5.2.2 Experiment Task and Procedure

Participants were tasked to write motivational messages and answer questions with the following procedure:
Read the introduction to describe the experiment objective and consent to the study.
1. Complete a 4-item word associativity test [19] to screen for English language skills.
2. Write 5 messages to motivate for physical activity for a fitness mobile app. For each message, one at a time,
   a) On the first page, depending on condition, see no prompt or a prompt with one or three phrases selected randomly or by Directed Diversity (see Figure 5), then write a motivational message in one to three sentences. This page is timed to measure *ideation task time*.
   b) Rate on a 5-point Likert scale the experience of ideating the current message: *ease of ideation* (described in Section 4.2.4), self-assessed success in writing *motivationally*, and success in writing *creatively* (Section 4.2.2); perception of the prompt on: *understandability*, *relevance to domain topic* (physical activity), *relevance to task* (motivation), *helpfulness* for inspiration, and *unexpectedness* (Section 4.2.3).
   c) Reflect and describe in free text on their rationale, thought process, phrase word usage, and ideation effort. We analyze these quotes to verify our understanding of the collected quantitative data.
3. Answer demographics questions, and end the survey by receiving a completion code.

---

[11] This is analogous to standard error is to standard deviation

### 5.2.3 Experiment Data Collection and Statistical Analyses

We recruited participants from Amazon Mechanical Turk with high qualification (≥5000 completed HITs with >97% approval rate). Of 282 workers who attempted the survey, 250 passed the test to complete the survey (88.7% pass rate). They were 45.2% female, between 21 and 70 years old (M=38.6); 76.4% of participants have used fitness apps. Participants were compensated after screening and were randomly assigned to one prompt selection technique. Participants in the None condition were compensated with US$1.80, while others with US$2.50 due to more time needed to answer the additional survey questions about prompts. Participants completed the survey in median time 15.4 minutes and were compensated >US$8/hour. We collected 5 messages per participant, 50 participants per condition, 250 ideations per condition, and 1,250 total ideations.

For all response variables, we fit linear mixed effects models described in Appendix Tables 20-23. To allow a 2-factor analysis, we divided responses in the None(0) condition (no prompt, 0 phrases) randomly and evenly to None(1) and None(3). Results are shown in Figure 6. We performed post-hoc contrast tests for specific differences identified. Due to the large number of comparisons in our analysis, we consider differences with p<.001 as significant and p<.005 as marginally significant. Most significant results reported are p<.0001. This is stricter than a Bonferroni correction for 50 comparisons (significance level = .05/50). We next describe the statistically significant results for prompt mediation check (RQ1.2), mediation analysis (RQ2.1, 2.2), and ideation evaluation (RQ3.1, 3.2). We include participant quotes from their rationale text response where available and relevant.

### 5.2.4 Results of Manipulation Check on Creativity and Mediation on Ideation Effort (RQ1.2, 2.1)

We discuss findings on how ideators perceived creativity factors in prompts and how prompt configurations affected their ideation effort. Figure 6 (Top) shows that compared to Random, Directed Diversity selected prompts that were more unexpected (good for diversity); but were slightly more difficult to understand (by half unit on 5-point Likert scale), very slightly less relevant (1/4 unit), and of slightly lower quality (1/2 unit). However, the relevance of the selected diverse prompts was not explicitly controlled. P173 in Directed(1) felt that the phrase "<u>first set of challenges is</u>" was *"straightforward and gave me the idea of what to write. It was very easy"*; whereas P157 in felt that the phrase "<u>review findings should be</u>" *"didn't really have anything I could think to tie towards a motivational message. I tried to think of it as looking back to see progress in terms of reviewing your journey."* Random prompts with more phrases were harder to understand, perhaps, because they were randomly grouped and are less semantically similar. P128 in Random(3) found that *"these* [phrases] *were hard to combine since they deal with different aspects of exercise. Also the weight lifting seems to be not the best thing for addressing obesity, so that was hard to work in."*

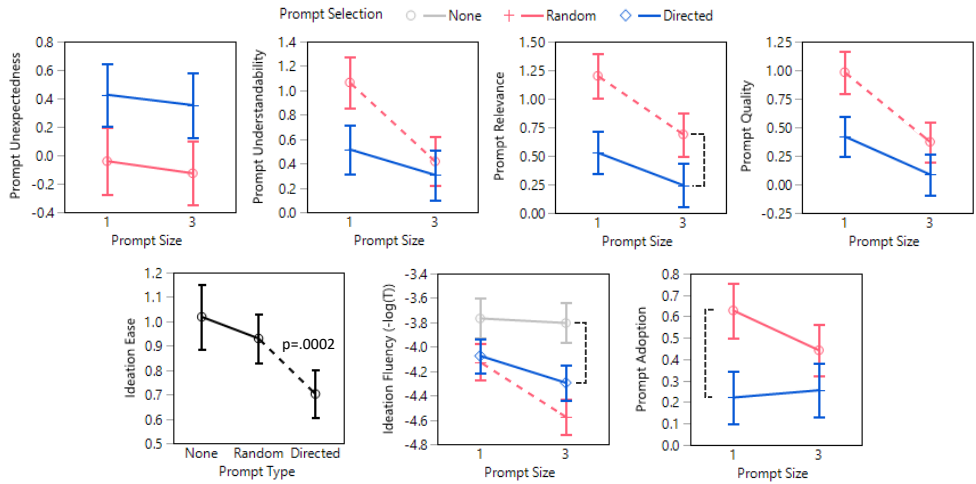

**Figure 6:** Results of ideators' perceived prompt creativity (Top) and ideation effort (Bottom) for different Prompt Selection technique and Prompt Size. All factors values on a 5-point Likert scale (−2="Strongly Disagree" to 2="Strongly Agree"). Dotted lines indicate extremely significant p<.0001 comparisons, otherwise very significant with p-value stated; solid lines indicate no significance at p>.01. Error bars indicate 90% confidence interval.

We found that ideation effort was mediated by prompt factors. Figure 6 (Bottom) shows that Directed prompts were least easy to use for ideation, and less adopted than Random selected prompts. This is consistent with Directed prompts being less understandable than Random. Ideating with 1-phrase prompts increased ideation time from 44.1s by 21.6s (48.9%) compared to None, and viewing 3 phrase increased time further by 11.9s. In summary, Directed

Diversity may improve diversity by selecting unexpected prompts, but at some cost of ideator effort and confusion. This cost compromises prompt adoption and suggests that directing diversity may not work. Yet, as we will show later, Directed Diversity does improve ideation creativity. We analyzed the confound of understandability further in Appendix Section K. Next, we investigate if prompts characteristics mediate more ideation creativity.

### 5.2.5 Results of Mediation Analysis of Diversity Propagation from Prompt to Ideation (RQ2.3)

We found that prompt configuration and perceived prompt creativity mediated the individual diversity of ideated messages (RQ2.2). Appendix Table 21a (in) shows that Ideation Mean (or Min) Pairwise Distance increased with Prompt Mean (or Min) Pairwise Distance by +0.176 (or +0.146), and marginally with Intra-Prompt Phrase Mean Distance by +0.021 (or +0.020). This means that farther Prompts stimulated farther Ideations, and higher variety of Phrases within each prompt drove slightly farther Ideations too. Hence, prompt diversity (mean pairwise distance) influenced ideation diversity, and prompt redundancy (minimum pairwise distance) influenced ideation redundancy. Appendix Table 21b shows that as Prompt Relevance decreased by one Likert unit (on 5-point scale), ideation mean pairwise distance decreased by 0.0034 (7.9% of ideation pairwise distance SD of 0.043) and ideation minimum pairwise distance decreases by 0.0056 (13% of SD). This suggests that prompting with irrelevant phrases slightly reduced diversity, since users had to have to conceive their own inspiration; e.g., P165 in Directed(1) *"couldn't make sense of the given messages, so I tried my best to make something somewhat motivational and correct from them."*. Prompt understandability and quality did not influence ideation individual diversity (p=n.s.). In summary, selecting and presenting computationally diverse and less redundant prompts increased the likelihood of crowdworkers ideating messages that are more computationally diverse and less redundant.

### 5.2.6 Results on Evaluating Individual, Collective Objective, Thematic Ideation Diversity (RQ3)

Having shown the mediating effects of diverse prompts, we now evaluate how prompt selection techniques affect self-assessed creativity ratings, objective diversity metrics of ideations, and thematically coded diversity metrics of ideations. To carefully distinguish between the commonly used mean pairwise distance with the less used minimum pairwise distance, we performed our analyses on them separately. We calculated one measurement of each collective diversity metric in Table 3 for all messages in each prompt selection condition, and computed uncertainty estimations from synthesized 50 bootstrap samples[12] to generate 50 readings of each diversity metric. We performed factor analyses on the metrics as described in Section 4.3, and performed statistical analyses on these factors as described in Appendix Table 22. Analyses on both individual diversity and collective diversity measures had congruent results (Figure 7), though results for collective diversity had more significant differences (p<.001). For collective diversity, our factor analysis found that Ideation Dispersion was most correlated with mean pairwise distance, and Ideation Evenness with entropy and mean of Chamfer distance. Directed(3) improved Ideation Dispersion from None, while Random reduced Dispersion (even more for 3 vs. 1 phrases). Directed prompts improved Ideation Evenness more than Random with respect to None. There was no significant difference for self-assessed Ideation Quality (p=n.s., Table 23a in Appendix).

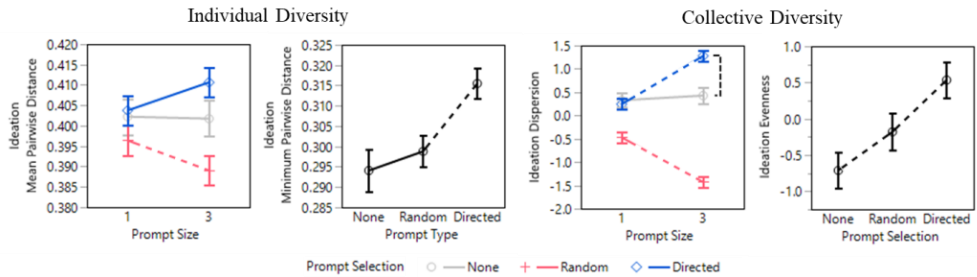

**Figure 7: Results of computed individual and collective diversity from ideations for different prompt configurations. See Figure 6 caption for how to interpret charts.**

The previous ideation diversity metrics were all computational. We next assess diversity with human judgement based on thematic analysis. To conserve manpower to evaluate ideations, we limited thematic coding and crowdworker validation to ideations from three conditions of prompts with 1 phrase, i.e., None, Random(1), and Directed(1). From the results of computational metrics, we expect bigger differences between Directed(3) and

---
[12] For each dataset, randomly sample with replacement from the original dataset until the same dataset size is reached.

Random(3) for this analysis too. From our thematic analysis, we coded 239 categories[13] which we consolidated to 53 themes (see Table 19 in Appendix). Figure 8 shows results from our statistical analysis. We found that ideations generated with Directed prompts had higher Flexibility and Originality in categories and themes than with Random or None. Ideations from Random prompts mostly had higher Flexibility and Originality compared to None, but the theme Originality was significantly lower. This could be because Random prompts primed ideators to fixate on fewer broad ideas (themes), instead of the higher number of fine-grained idea categories.

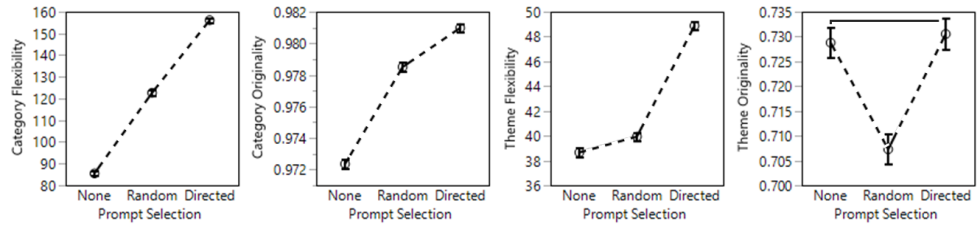

Figure 8: Results of diversity in categories and themes derived from thematic analysis of ideations.

In summary, despite lower ideation ease and understandability with Directed prompts (Section 5.2.4), we found objective and thematic evidence that Directed Diversity improved ideation diversity compared to Random and None. Next, we describe how crowdworkers would rate these ideations.

## 5.3 Validation User Studies

The third and fourth studies employed third-party crowdworkers to assess the creativity of ideated messages from the Ideation User Study, to answer *(RQ3) How do prompt selection techniques affect diversity in ideations?* This provides a less biased validation than asking ideators to self-assess. We conducted three experiments with different questioning format to strengthen the experiment design. Appendix Figures 20-24 details the questionnaires.

### 5.3.1 Individual Validation: Experiment Treatment and Procedure

For the individual validation study, we conducted a within-subjects experiment with prompt selection technique (None, Random, Directed) as independent variable, and controlled prompt size ($g = 1$). Each participant assessed 25 ideation messages chosen randomly from the three conditions. Participants went through the same procedure as in the Ideation user study, but with a different task in step 3:

3. Assess 25 messages regarding how well they motivate for physical activity. For each message,
   a) Read a randomly chosen message.
   b) Rate on a 7-point Likert scale, whether the message is *motivating* (effective), *informative*, and *helpful* (as described in Section 4.2.4).
   c) Reflect and write the rationale in free text on why they rated the message as effective or ineffective. This was only asked randomly two out of 25 times, to avoid fatigue.

As we discuss later, we found that participants confounded the three ratings questions and answered them very similarly (responses were highly correlated), thus, we designed collective validation user studies to pose different questions and distinguish between the measures.

### 5.3.2 Collective Ranking Validation: Experiment Treatment and Procedure

The collective validation study had the same experiment design as before, but different procedure step 3:

3. Complete 5 trials to rate collections of ideation messages, where for each trial,
   a) Study three groups of 5 messages each (3×5 messages) to
   b) Rank message groups as most, middle or least *motivating*, *informative*, and *unrepetitive* (Section 4.2.4).

Instead of rating messages individually, participants viewed grouped messages from each condition side-by-side and answered ranking questions. Messages in each group were selected from those ideated with the same prompt selection technique. By asking participants to assess collections rather than individual messages, we explicitly measured perceived diversity, since the user perceived the differences between all ideations in the collection; this is more direct than asking them about the "informativeness" of an ideation, since this could be confounded with "helpfulness", "teaching something new", "telling something different from other messages", etc. This approach differs from the triplet similarity comparison [55,91] employed by Siangliulue et al. [79], and benefits from requiring fewer assessments. We asked participants to rank groups rather than rate them relatively to obtain a forced choice [25]. Another method to assess diversity involves longitudinal exposure (e.g., [50]), but this is expensive and difficult to scale.

---
[13] Example categories (in themes): Pull-ups (Exercise Suggestion), Strong immune system (Health Benefits), Set daily exercise goal (Goals). See Appendix 8.7 for full list of categories and themes.

### 5.3.3 Collective Pairwise Rating Validation: Experiment Treatment and Procedure

The collective pairwise rating validation study further validates our results with an existing, commonly used measure to rate the difference between pairs of messages, both from the same prompt selection technique [27,79]. We randomly selected 200 message-pairs from None, Random(1) and Directed(1), yielding a pool of 600 message-pairs. All steps in the procedure are identical as before except for Step 3:

3. Rate 30 message-pairs randomly selected from the message-pair pool, where for each message-pair,
   a) Read the two messages
   b) Rate their difference on a 7-point Likert scale: 1 "Not at all different (identical)" to 7 "Very different"

This complements the previous study by having participants focus on two messages to compare, which is more manageable than assessing 5 messages, but is limited to a less holistic impression on multiple messages.

### 5.3.4 Experiments Data Collection and Statistical Analysis

For all validation studies, we recruited participants from Amazon Mechanical Turk with the same high qualification as the ideation study. Of 348 workers who attempted the surveys, 290 passed the screening tests to complete the surveys (83.3% pass rate). They were 50.2% female, between 22 and 71 years old (M=38.1); 67.5% of participants have use fitness apps. For the individual validation study, Participants completed the survey in median time 14.7 minutes and were compensated US$1.50; for the collective ranking validation study, participants completed the survey in median time 12.7 minutes and were compensated US$1.80; for the collective pairwise rating validation study, participants completed the survey in median time 8.4 minutes and were compensated US$1.00. In total, 740 messages were individually rated 3,375 times (M=4.56x per message), 450 message groups were ranked 1,350 times (M=3.00x per message group), and 600 message pairs were rated 2,430 times (M=4.05x per message pair). To assess inter-rater agreement, we calculated the average aggregate-judge correlations [18] as r=.59, .62, .63 for motivation, informativeness and helpfulness for individual validation ratings, respectively; these were comparable to Chan et al.'s r=.64 for idea novelty [18].

We performed the same statistical analyses as in the Ideation User Study (see Section 5.2.3), report the linear mixed effects models in Appendix Table 23, and include participant quotes from their rationale text response where relevant. For the collective ranking validation study, we counted how often each Prompt Selection technique was ranked first or last across the 5 trials, performed factor analyses on the counts for best and worst ranks for the three metrics (*motivating*, *informative*, *unrepetitive*) to derive three orthogonal factors (Ideations Unrepetitive, Ideations Informative, Ideations Motivating), and performed the statistical analysis on the factors (see Table 23b in Appendix).

### 5.3.5 Results on Evaluating Individual and Collective Ideation Creativity (RQ3)

We investigated whether Directed prompts stimulate the highest ideation diversity and whether 3rd-party validations agree with our computed and thematic results. For illustration, Appendix Table 25 shows examples of message-groups with high and low factor values.

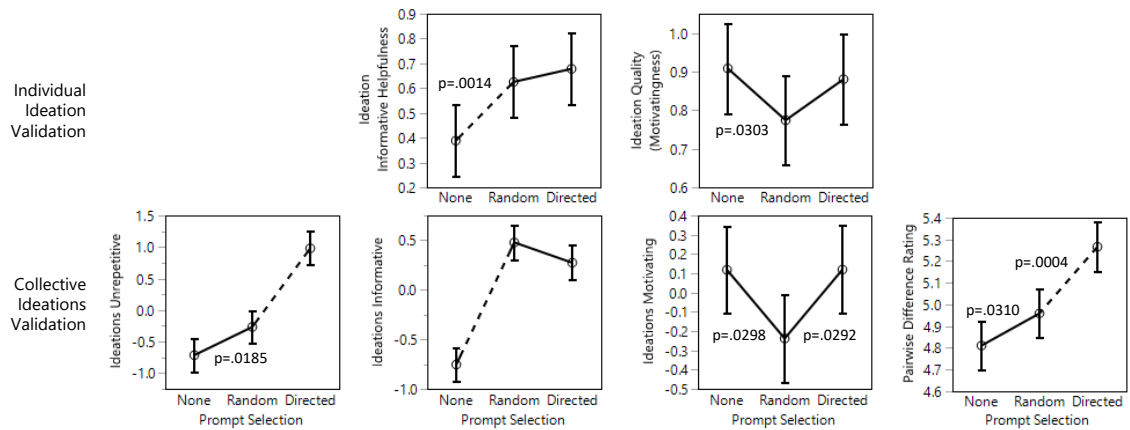

Figure 9: Results of perceived individual and collective creativity from the three validation user studies.

Figure 9 shows results of our statistical analysis. We found that ideations from Directed prompts were most different and least repetitive, ideations from Random were no different and as repetitive as None. Ideations generated with prompts were more informative and helpful than without prompts, but there was no difference whether the prompts were Directed or Random. For example, P4 reviewed the message "Exercise and live longer, and prosper more!" ideated with None, and felt that *"it's basically telling you what you already know. It's a rather generic message."*; P63 reviewed the message "Waking up early and working out will help you get into shape, and is a great way to have more energy and better sleep." from the Directed(1) prompt "into a habit of sleep" and felt *"it's effective because it gives me a goal and tells me why this is a good goal"*. There were no significant differences in ideation quality or motivation, though there was a marginal effect that Random prompts could hurt quality compared to None. Therefore, Directed Diversity helped to reduce ideation redundancy compared to Randomly selected prompts, improved informativeness, and did not compromise quality.

### 5.4 Summary of Answers to Research Questions

We summarize our findings to answer our research questions with results from multiple experiments.

**RQ1.** *How did prompt selection techniques affect diversity in prompts?* Compared to Random, Directed Diversity: a) selected more diverse prompts, b) with less redundancy from prior ideation, c) that ideators perceived as more unexpectedness, but d) of poorer quality and understandability.

**RQ2.** *How did diversity in prompts affect the ideation process for ideators?* Compared to Random, prompts selected with Directed Diversity were: a) harder to ideate with, b) less applied for ideation, c) but their higher prompt diversity somewhat drove higher ideation diversity.

**RQ3.** *How did prompt selection techniques affect diversity in ideations?* Compared to None and Random, Directed Diversity: a) improved ideation diversity and reduced redundancy, b) increased the flexibility and originality of ideated categories, c) without compromising ideation quality.

## 6 Discussion

We discuss the generalization of our technical approach, evaluation framework, and experiment findings.

### 6.1 Need for Sensitive and Mechanistic Measures of Creativity

We have developed an extensive evaluation framework for two key reasons: 1) to precisely detect effects on diversity, and 2) to track the mechanism of diversity prompting. We have sought to be very diverse in our evaluation of prompt technique to carefully identify any benefits or issues. We have found that some popular metrics (e.g., mean pairwise distance) were less sensitive than others (e.g., MST Dispersion / Remote-tree). Therefore, a null result in one metric (e.g., [79]) may not mean that diversity was not changed (if measured by another metric). Instead of only depending on the "black box" experimentation of prompt treatment on ideation (e.g., [18,40,79,80]), investigating along the ideation chain is interpretable and helpful for us to identify potential issues or breakdowns in the diversity prompting mechanism. Had our evaluation results on ideation diversity been non-significant, this would be helpful to debug the lack of effectiveness.

Conversely, we may find that an ideation diversity effect may be due to contradictory or confounding effects. Indeed, we found that Directed Diversity improved diversity, despite poorer prompt understandability and adoption. Ideators could not directly use the selected prompts, but still managed to conceive ideas that were more diverse than not having seen prompts or seeing random ones. This suggests that they generated ideas sufficiently near the prompts. The findings also suggested that the increased effort helped to improve diverse ideation [5,6,96], but the ideator user experience should be improved. Future work is needed to improve Directed Diversity to reduce ideator effort and improve the relevance of selected prompts, such as by limiting the distance of new prompts from prior ideations, or using idea-based embeddings [79,80] instead of language models, as discussed next.

### 6.2 Generalization of Directed Diversity to other Domains

The full process of Directed Diversity (Figure 1) allows us to generalize its usage to other domains, such as text creativity tasks beyond motivational messages (e.g., birthday greetings [79]) by changing the document sources in the phrase extraction step. In the phrase embedding step, we used the Universal Sentence Encoder [14], but other text embedding models (e.g., word2vec [63], GloVe [71], ELMo [74], BERT [23]) could be used that model languages slightly differently. In the third step, we selected phrases based on the Remote-tree diversity formulation using an efficient greedy algorithm that approximates the diversity maximization. Other diversity criteria and maximization algorithms could be used (see review [20]). Note that since USE and similar language models are domain-independent, which do not model the semantics of specific domains and semantic quality, Directed Diversity cannot guarantee improving quality. A domain-specific model trained with human-annotated labels of quality could be used to improve both diversity and quality. Furthermore, instead of representing text with language models, the idea space could be explicitly modelled to obtain embeddings from annotated semantic similarity [55,79]. Finally, since Directed Diversity operates on a vector representation of prompting and ideations, it can also be used for ideation tasks beyond text as long as they can be represented in a feature vector by feature engineering or with deep learning approaches, such as furniture [58], mood boards [49], and emojis [101].

## 6.3 Generalization of Evaluation Framework

Our Evaluation Framework is a first step towards the goal of standardizing the evaluation of crowd ideation. This requires further validation and demonstration on existing methods of supporting crowd ideation. Due to the costs of engineering effort, set-up preparation, and recruitment, we defer it to future work. Just as the Directed Diversity pipeline is generalizable, we discuss how the Diversity Prompt Evaluation Framework is generalizable. We had identified many diversity metrics, but only measured some of them; see [20] for a review of other mathematical metrics. If applying the framework to non-text domains, the vector-based distance metrics should still be usable if the concepts can be embedded with a domain model. While we analyzed diversity in terms of mathematical metrics [20] and several measures for creativity [92], other criteria may be important to optimize, such as serendipity for recommender systems to avoid boredom [44].

To measure creativity, just as in prior research [50], we had used several Likert scale ratings (e.g., helpfulness and informativeness) and found evidence that participants confound them. Furthermore, it may be excessive to apply all our measures, therefore the researcher is advised to use them judiciously. For example, we found that individually rating ideations tends to lead to poor statistical significance, so this data collection method should be avoided. The thematic analysis coding is also very labor intensive for the research team, but provides rich insights into the ideas generated. We had proposed using ranking and pairwise rating validations of collections of ideations as a scalable way to measure collective diversity.

While our evaluations based on generating motivational messaging for physical activity helped to provide a realistic context, it was limited to measuring preliminary impressions of validators. The social desirability effect may have limited how accurately participants rated the effectiveness of the messages. While our focus was on evaluating diversity, future work that also seeks to improve and evaluate motivation towards behavior change should conduct longitudinal trials with stronger ecological validity [50].

## 7 Conclusion

In this paper, we presented Directed Diversity to direct ideators to generate more collectively creative ideas. This is a generalizable pipeline to extract prompts, embed prompts using a language model, and select maximally diverse prompts. We further proposed a generalizable Diversity Prompting Evaluation Framework to sensitively evaluate how Directed Diversity improves ideation diversity along the ideation chain — prompt selection, prompt creativity, prompt-ideation mediation, and ideation creativity. We found that Directed Diversity improved collective ideation diversity and reduce redundancy. With the generalizable prompt selection mechanism and evaluation framework, our work provides a basis for further development and evaluations of prompt diversity mechanisms.

## 8 Acknowledgements

This work was carried out in part at NUS Institute for Health Innovation and Technology (iHealthtech) and with funding support from the NUS ODPRT and Ministry of Education, Singapore.

# A Definitions of Prompt Selection Variables

Table 6: Independent variables used in the simulation and user studies to manipulate how prompts are shown to ideators.

| Variable | Definition | Interpretation |
| --- | --- | --- |
| Prompt Selection | *None*: no prompt, other than task instructions<br>*Random*: randomly selected phrase from corpus<br>*Directed*: prioritized phrase from corpus | Selection algorithm for selecting phrases to include in prompts. |
| Prompt Count | Number of prompts<br>$\{50, 100, 150, 200, 250, \ldots, n_{prompts}\}$ | Indicates how many prompts shown to generate new messages. A prompt may contain $\geq 1$ phrases. This was only tested in the simulation study. |
| Prompt Size | Number of phrases per prompt<br>$\{1,2,3,4,5\}$ | Prompts selected depends on Prompt Selection. |

# B Additional Definitions of Diversity Metrics

## B.1 Thematic Analysis Method for Flexibility and Originality Metrics

*Flexibility* [85] measures how many unique ideas (conceptual categories) was generated, and *originality* [100] measures how infrequently each conceptual category occurs. These require expert annotation to identify distinct categories. We conducted a thematic analysis of ideated messages using open coding of grounded theory [34] to derive categories. These categories were added, reduced, merged, and refined by iteratively assessing the messages. We then consolidated the categories into themes using affinity diagramming [8]. This was done separately for different prompt techniques. The thematic analysis was primarily performed by one co-author researcher with regular discussion with co-authors who are experienced HCI researchers with experience in Amazon Mechanical Turk experiments and research on health behavior change. We calculated inter-rater reliability on a random 10% subset of messages was coded independently by another co-author to obtain a Krippendorff's alpha with MASI distance [70] of $\alpha = 0.82$, which indicated good agreement. Note that while thematic analyses and affinity diagramming are popular methods to interpret qualitative data, we use them here for data pre-processing. Finally, we calculate the flexibility and originality measures based on the coded categories (fine-grained) and themes (coarser) described in Table 7.

Table 7: Metrics of creativity of ideation based on categories and themes derived from a thematic analysis of generated ideas. Metrics are shown for categories, but are the same for themes.

| Metric | Definition | Interpretation |
| --- | --- | --- |
| *Messages Flexibility* | Number of categories coded<br>$\sum_c [f_c > 0]$ | This counts how many unique categories/themes were observed in messages for each Prompt Technique. A higher count indicates qualitatively more diversity. |
| *Messages Originality* | Category originality<br>$o^c = (1 - f_c/N_p)$ | How original $o^c$ each theme is, where $f_c$ is the frequency for the category $c$, $N_p$ is the number of messages with the Prompt Technique $p$. |

## B.2 Intra-Prompt Diversity Metrics based on Embedding Distances

Table 8: Metrics of prompt diversity for all phrases in a single prompt.

| Metric | Definition | Interpretation |
| --- | --- | --- |
| *Intra-Prompt Mean Phrase Distance* | Intra-prompt mean phrase-phrase distance<br>$\frac{1}{g} \sum_{i,j \in Prompt} d(x_i^P, x_j^P)$ | Indicates how similar (consistent) all phrases are to one another in the same prompt. Prompts with better consistency would be easier to understand and use. |
| *Prompt Phrase Chamfer distance* | $\frac{1}{g} \sum_{i \in Prompt} \min_{j \notin i} d(x_i^P, x_j^P)$ | Average distinctiveness of phrases in prompt. |

## C Definitions of Prompt Adoption Metrics

Table 9: Metrics indicating how much of prompt text and concepts are adopted into the ideations.

| Metric | Definition | Interpretation |
|---|---|---|
| *Prompt Recall* | $\frac{1}{g}\sum_{Phrase \in Prompt} \frac{n_{word \in Ideation \wedge word \in Phrase}}{n_{word \in Phrase}}$ | The proportion of words from phrases that were used in the ideated message. |
| *Prompt Precision* | $\sum_{phrase \in prompt} \frac{n_{word \in Ideation \wedge word \in Phrase}}{n_{word \in Ideation}}$ | The proportion of ideated message words that were from phrases in the shown prompt. |
| *Prompt-Ideation Distance* | Prompt-Ideation distance $d(x_i^{Pr}, x_j^{I})$ | Indicates how similar the written ideation message is to the prompt, as a measure of how the phrase(s) ideas were adopted. |

## D Pairwise Embedding Distances of Phrases and Messages

These figures show the distribution of pairwise distances based on the embeddings of phrases and messages.

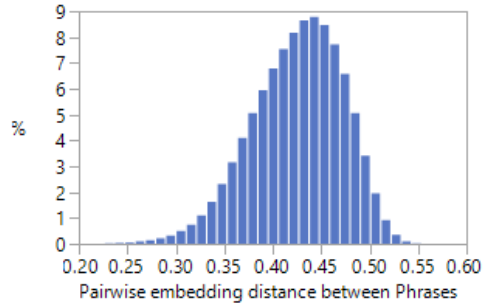

Figure 10: Distribution of pairwise distances between the extracted phrases (N=3,666). The pairwise distances ranged from Min=0.057 to Max=0.586, Median=0.430, inter-quartile range 0.394 to 0.460, SD= 0.047.

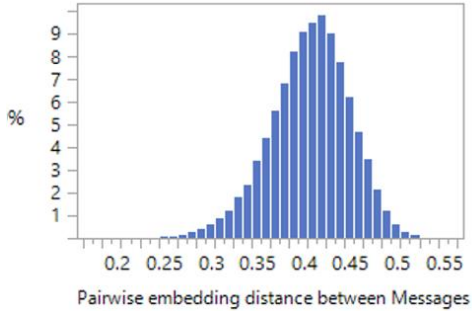

Figure 11: Distribution of pairwise distances between the messages (N=250) ideated in the pilot study with no prompting (None). The pairwise distances ranged from Min=0.169 to Max=0.549, Median=0.405, inter-quartile range 0.376 to 0.432, SD=0.043.

# E Results of Characterization Simulation Study

We created 50 simulations for each prompts configuration to get a statistical estimate of the performance of each prompt selection technique. Figure 12 shows the results from the simulation study. Error bars are extremely small, and not shown for simplicity. Span and Sparseness results not shown, but are similar to Mean Distance. Note that we computed the mean of MST edge distances instead of sum, which is independent of number of prompts. In general, Directed Diversity selects prompts to be more diverse for fewer prompts (smaller prompt count), but after a threshold, Random selection can provide for better diversity. This demonstrates directing is useful for small crowd budgets. Note that the actual threshold depends on corpus and application domain. We found an interaction effect where single-phrase prompts benefit most with Directed Diversity, since for low prompt count, and Directed(1-phrase) has highest diversity, followed by Directed(3), Random(3), and Random(1) with lowest diversity.

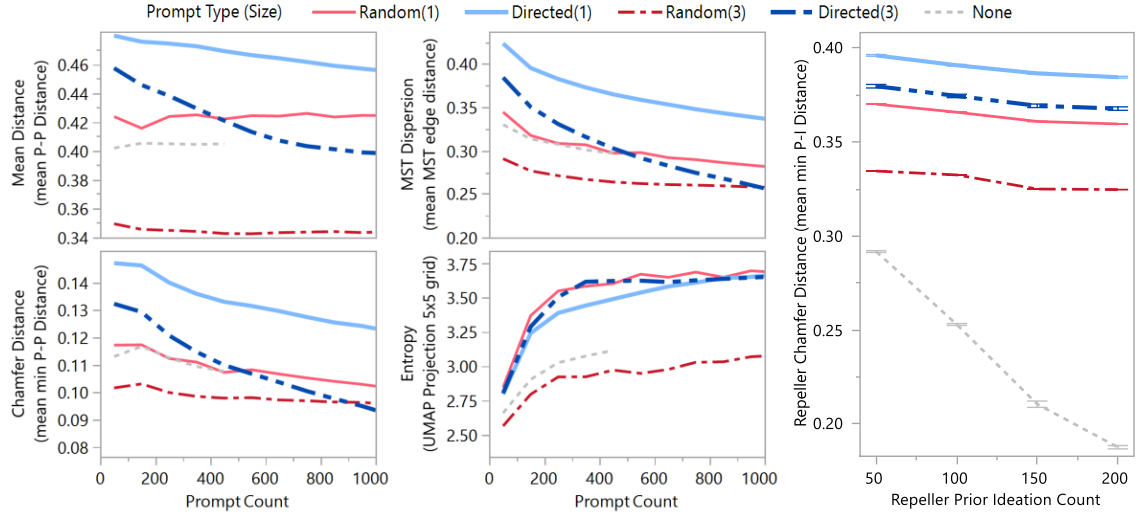

Figure 12: Influence of prompt selection technique, prompt size, and prompt count on various distance and diversity metrics. Higher values for all metrics indicate higher diversity. Span and Sparseness results are not shown, but are similar to Mean Distance. Note that we computed the mean of MST edge distances instead of sum, which is independent of number of prompts. Error bars are extremely small, and not shown for simplicity.

# F  Factor Loadings from Factor Analysis in User Studies

Table 10: The rotated factor loading of factor analysis on metrics of prompt distance and consistency. Factors explained 73.6% of the total variance. Bartlett's Test for Sphericity to indicate common factors was significant ($\chi^2$= 5810, p<.0001).

|  | Prompt Distance | Prompt Consistency |
|---|---|---|
| Phrase Minimum Pairwise Distance | 0.95 | 0.08 |
| Prompt Minimum Pairwise Distance | 0.95 | 0.05 |
| Intra-Prompt Mean Phrase Distance | -0.05 | -0.69 |

Table 11: The rotated factor loading of factor analysis on metrics of perceived helpfulness of prompts. Factors explained 68.9% of the total variance. Bartlett's Test for Sphericity to indicate common factors was significant ($\chi^2$= 2575, p<.0001).

|  | Prompt Quality | Prompt Unexpectedness | Prompt Relevance | Prompt Understandability |
|---|---|---|---|---|
| Phrase Helpfulness rating | 0.85 | -0.13 | 0.2 | 0.15 |
| Phrase Relevance to Task (Motivation) rating | 0.89 | -0.2 | 0.23 | 0.14 |
| Phrase Understanding rating | 0.62 | -0.21 | 0.27 | 0.47 |
| Phrase Relevance to Domain (Exercise) rating | 0.59 | -0.14 | 0.54 | 0.18 |
| Phrase Unexpectedness rating | -0.11 | 0.63 | -0.06 | -0.05 |

Table 12: The rotated factor loading of factor analysis on metrics of prompt adoption. Factors explained 65.1% of the total variance. Bartlett's Test for Sphericity to indicate common factors was significant ($\chi^2$= 1315, p<.0001).

|  | Phrase Adoption |
|---|---|
| Prompt Precision | 0.82 |
| Prompt Recall | 0.67 |
| Prompt-Ideation Distance | -0.92 |

Table 13: The rotated factor loading of factor analysis on diversity metrics of generated messages. Factors explained 75.2% of the total variance. Bartlett's Test for Sphericity to indicate common factors was significant ($\chi^2$= 2676, p<.0001).

|  | Ideation Dispersion | Ideation Evenness |
|---|---|---|
| Message Remote-clique | 0.99 | 0.16 |
| Message Sparseness | 0.99 | 0.16 |
| Message Span | 0.77 | -0.05 |
| Message MST Dispersion | 0.29 | 0.96 |
| Message Chamfer Distance | -0.02 | 0.91 |
| Message Entropy | 0.01 | 0.3 |

Table 14: The rotated factor loading of factor analysis on metrics of perceived quality of the generated messages. Factors explained 80.9% of the total variance. Bartlett's Test for Sphericity to indicate common factors was significant ($\chi^2$= 5810, p<.0001).

|  | Ideation Informative-Helpfulness | Ideation Quality |
|---|---|---|
| Informativeness rating | 0.8 | 0.39 |
| Helpfulness rating | 0.66 | 0.65 |
| Motivation rating | 0.39 | 0.79 |

Table 15: The rotated factor loading of factor analysis on metrics of group ranking of the generated messages. Factors explained 93.9% of the total variance. Bartlett's Test for Sphericity to indicate common factors was significant ($\chi^2$= 366, p<.0001). For usability, "unrepetitive" was measured with the word "repetitive" in the survey.

|  | Ideations Unrepetitive | Ideations Informative | Ideations Motivating |
|---|---|---|---|
| Sum(Most Unrepetitive (Rank=1)) | 1.32 | 0.40 | 0.10 |
| Sum(Most Informative (Rank=1)) | 0.48 | 0.72 | 0.05 |
| Sum(Least Unrepetitive (Rank=3)) | -0.73 | -0.50 | 0.02 |
| Sum(Least Informative (Rank=3)) | -0.26 | -0.89 | -0.03 |
| Sum(Most Motivating (Rank=1)) | 0.17 | -0.04 | 1.00 |
| Sum(Least Motivating (Rank=3)) | 0.05 | -0.06 | -0.54 |

Table 16: The rotated factor loading of factor analysis on metrics of message distinctness. Factors explained 74.8% of the total variance. Bartlett's Test for Sphericity to indicate common factors was significant ($\chi^2$= 1022, p<.0001).

|  | Ideation Distance |
|---|---|
| Ideation Min Pairwise Distance | 0.86 |
| Ideation Mean Pairwise Distance | 0.86 |

Table 17: The rotated factor loading of factor analysis on metrics of ideation effort. Factors explained 59.0% of the total variance. Bartlett's Test for Sphericity to indicate common factors was significant ($\chi^2$= 1008, p<.0001).

|  | Ideation Self-Quality | Ideation Ease |
|---|---|---|
| Message Creativity Self-Rating | 0.76 | 0.19 |
| Message Motivation Self-Rating | 0.63 | 0.55 |
| Message Writing Ease | 0.76 | 0.19 |

## G  Survey Screenshots in User Studies

## G.1  Ideation User Study

**Task Instructions:**

You will now write 5 messages to encourage someone to exercise.

Please follow these guidelines:

- Messages should generally be one to three sentences long
- Write **concise** messages (imagine the message could be sent to people on a smartphone)

Please do not consult online sources for your answers, and generate messages yourself to the best of your ability.

HITs with poor levels of English or repeated answers may be rejected.

Figure 13: The instructions in the Ideation User Study for the None condition.

Write a message to **motivate** someone to exercise.

Messages should be **concise** and **one to three sentences long**.

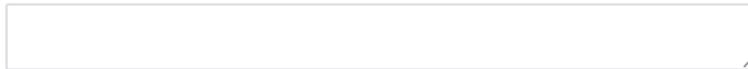

Figure 14: For the None, users are asked to write a message that is at least one to three sentences long.

Task Instructions:

You will now write 5 messages to **motivate** someone to exercise, and **you will be shown short phrases to inspire you**.

Please follow these guidelines:

- Use each phrase for inspiration to write a message that would **motivate** someone to exercise.
- You do not need to use the phrases word for word (although you can if you want to).
- You can write using the high-level theme of a phrase (e.g. a phrase might be about weightlifting).
- While it's good to use the phrases, do not force the use of a phrase if it would not produce a **motivating** or **grammatically correct** message. Just try to interpret a theme from each phrase if you can.
- All phrases are machine generated, and may not be perfect. Please **ignore phrases that you don't find relevant** to writing motivational messages for exercise (but feel free to be creative if you can use any of the phrases to write your message).
- Messages should generally be 1 to 3 sentences long.
- Write **concise** messages (imagine the message could be sent people on a smartphone).

Please do not consult online sources for your answers, and generate messages yourself to the best of your ability.

While we try to avoid rejecting work, HITs that copy-paste the same message multiple times or write messages in *very* poor English may be rejected.

Figure 15: The instructions of Ideation User Study for the Random(1) and Directed(1) conditions.

Write a message to **motivate** someone to exercise.

Messages should be **concise** and **one to three sentences long**.

Please use the phrase below for inspiration:

"exercise can sometimes be"

Please double-check that your message is as **motivational** as you can write for this phrase, and do not use the phrase word for word if it would not produce a **motivating and grammatically correct** message.

Figure 16: Random(1) and Directed(1) prompts consisted of one phrase per prompt. Note that selected phrase for each trial will be different.

Task Instructions:

You will now write 5 messages to **motivate** someone to exercise, and **you will be shown short phrases to inspire you**.

Please follow these guidelines:

- Use the phrases for inspiration to write a message that would **motivate** someone to exercise.
- While it's good to use as many of the phrases as you can, do not force the use of the phrases if it would not produce a motivating message.
- Messages should generally be 1 to 3 sentences long.
- Write **concise** messages (imagine the message could be sent people on a smartphone).
- You can **summarise phrases with similar meanings**:
  - For example, if you were given the phrases:
    "Weight lifting is good", "lift weights to boost", "when lifting you will"
    You could write the message:
    "Weight lifting is good to boost your mood!"
- All phrases are machine generated, and may not be perfect. Please **ignore phrases that you don't find relevant** to writing motivational messages for exercise (but feel free to be creative if you can use any of the phrases to write your message)
- You can write using the entire phrases or fragments of the phrases.

Please do not consult online sources for your answers, and generate messages yourself to the best of your ability.

While we try to avoid rejecting work, HITs that copy-paste the same message multiple times or write messages in *very* poor English may be rejected.

Figure 17: The instructions of Ideation User Study for the Random(3) and Directed(3) conditions.

Write a message to **motivate** someone to exercise.

Messages should be **concise** and **one to three sentences long**.

Please use the phrases below for inspiration:

Phrase 1: "did not regularly exercise"
Phrase 2: "workout program would put"
Phrase 3: "strong correlation between exercise"

Please double-check that your message is as **motivational** as you can write for these phrases, and do not force the use of the phrases if it would not produce a **motivating and grammatically correct** message.

Figure 18: Random(3) and Directed(3) prompts consist of three phrases per prompt. Note that selected phrases for each trial will be different.

You were shown the phrases:

Phrase 1: "did not regularly exercise"
Phrase 2: "workout program would put"
Phrase 3: "strong correlation between exercise"

and wrote the message:

"If you find it difficult to regularly exercise, then why not try a workout program?"

Do you agree or disagree that...

|  | Strongly Disagree | Disagree | Neither | Agree | Strongly Agree |
|---|---|---|---|---|---|
| the message was **easy to write** | ○ | ○ | ○ | ○ | ○ |
| you successfully wrote a **motivating** message | ○ | ○ | ○ | ○ | ○ |
| you successfully wrote a **creative** message | ○ | ○ | ○ | ○ | ○ |

**For each phrase**, do you agree or disagree that it was…

|  | Phrase 1: "did not regularly exercise" | | | | | Phrase 2: "workout program would put" | | | | | Phrase 3: "strong correlation between exercise" | | | | |
|---|---|---|---|---|---|---|---|---|---|---|---|---|---|---|---|
|  | Strongly Disagree | Disagree | Neither | Agree | Strongly Agree | Strongly Disagree | Disagree | Neither | Agree | Strongly Agree | Strongly Disagree | Disagree | Neither | Agree | Strongly Agree |
| Helpful for inspiration | ○ | ○ | ○ | ○ | ○ | ○ | ○ | ○ | ○ | ○ | ○ | ○ | ○ | ○ | ○ |
| Easy to understand | ○ | ○ | ○ | ○ | ○ | ○ | ○ | ○ | ○ | ○ | ○ | ○ | ○ | ○ | ○ |
| Relevant to motivation | ○ | ○ | ○ | ○ | ○ | ○ | ○ | ○ | ○ | ○ | ○ | ○ | ○ | ○ | ○ |
| Relevant to physical exercise | ○ | ○ | ○ | ○ | ○ | ○ | ○ | ○ | ○ | ○ | ○ | ○ | ○ | ○ | ○ |
| Unexpected | ○ | ○ | ○ | ○ | ○ | ○ | ○ | ○ | ○ | ○ | ○ | ○ | ○ | ○ | ○ |

Please reflect and describe:

- your thought process when writing the message
- how you used the phrases
- what you found difficult or easy when writing your message.

Figure 19: Ideators are asked to evaluate the message they wrote by providing Likert scale ratings for many different factors along with a short reflection about the message writing process. The screenshot above shows the evaluation screen for Directed(3).

## G.2 Validation User Studies

Rating messages to encourage exercise:

Hello! Thank you for your interest in our study.

We'd like you to **rate 25 messages intended to encourage physical activity**. You will also be asked for **written feedback on two of the 25 messages**.

The survey should take less than **10 minutes**, and your responses will be anonymous.

Figure 20: The instruction for individual message rating tasks.

For the message below:

"Set a goal to indulge in something that keeps you more fit, like a new set of golf clubs!"

How **motivating or demotivating** do you personally find this message?

[ Very Demotivating ] [ Demotivating ] [ Somewhat Demotivating ] [ Neither ] [ Somewhat Motivating ] [ Motivating ] [ Very Motivating ]

How **informative or uninformative** do you personally find this message?

[ Very Uninformative ] [ Uninformative ] [ Somewhat Uninformative ] [ Neither ] [ Somewhat Informative ] [ Informative ] [ Very Informative ]

How **helpful or unhelpful** do you personally find this message?

[ Very Unhelpful ] [ Unhelpful ] [ Somewhat Unhelpful ] [ Neither ] [ Somewhat Helpful ] [ Helpful ] [ Very Helpful ]

Given the message is intended to encourage physical activity, **what do you think is effective or ineffective** about the message?

[You will only be asked for written feedback on two of the 25 messages you rate. Please take time to consider your answer and write out your thoughts in full.]

Figure 21: Validators rated a randomly selected message on a Likert scale and gave a justification.

Which group of messages is "best" at encouraging physical activity?

Hello! Thank you for your interest in our study.

We'd like you to compare different groups of messages intended to encourage physical activity.

The survey should take roughly **15 minutes**, and your responses will be anonymous.

Figure 22: The instruction for group message ranking tasks.

Please read the 3 groups of messages below.

For each group, imagine you receive one of the messages a day for 5 days in a row.

**Group 1:**

| Exercise isn't only good for the body, but it's also good for the mind. | There are plenty of ways to exercise, and you don't have to do it alone. | Small changes can make a big impact. Start slow and work your way towards your goals. Every step is an accomplishment! | Exercising will help you live longer. | That former Beefcake Body is still inside you, just waiting to come back out! Start slow and watch the progress. |

**Group 2:**

| Biceps may get all the love - but they'd be nothing without your triceps! Triceps workouts are quick, fun, and give your arms good balance. | Time for those 5 push ups. You can do it! | Many minutes of exercise should improve your mental well-being. At least 20 minutes can clear your head and feed your starved body the oxygen it needs to complete your other work. | Time to start a weight loss workout, you can do it! | Make a timetable to do vigorous exercise with rest periods in between. |

**Group 3:**

| Make sure you stretch to stay injury free as you exercise. You can do it! | Not all exercise needs to be for a long stretch of time. Make some time today! | Live healthy in the gym, and in the kitchen! You can't make everything from cheese dip, but you can still choose healthier alternatives, like cauliflower for example! Small changes today can lead to lasting results. | Sometimes an iPod is helpful for keeping your mind occupied while doing some exercises that may be repetitive. Get in the zone! | Get in the right mind set & you too will become one of the envied - those for whom exercise is second nature. All it takes is a shift in attitude. You can do whatever you put your mind to. Know it. Live it. Start today. |

Please order how **motivating** you find each group of messages.

[Drag and drop the groups to add your answer.]

| Items | Most Motivating Group | Second Most Motivating Group | Least Motivating Group |
|---|---|---|---|
| Group 1 | | | |
| Group 2 | | | |
| Group 3 | | | |

Please order how **informative** you find each group of messages.

| Items | Most Informative Group | Second Most Informative Group | Least Informative Group |
|---|---|---|---|
| Group 1 | | | |
| Group 2 | | | |
| Group 3 | | | |

Please rank how **repetitive** you find each group of messages.

| Items | Most Repetitive Group | Second Most Repetitive Group | Least Repetitive Group |
|---|---|---|---|
| Group 1 | | | |
| Group 2 | | | |
| Group 3 | | | |

Figure 23: Validators were asked to rank groups of messages for motivation, informativeness and repetitiveness. Note that while we used the word "repetitive" for usability in the survey, we analyzed this dependent variable as "unrepetitive" to be consistent with other diversity metrics.

How different are these two messages?
(Hint: Judge their similarity or difference based on words and meaning.)

**Message 1:** Exercising can make you happier. Run for joy!

**Message 2:** Exercising is like a barrier against insecurity. You will feel so much better about yourself when you really explore what your body can do.

| 1 - Not at all different (identical) | 2 | 3 | 4 | 5 | 6 | 7 - Very Different |

Figure 24: Validators were asked to rate the difference of two messages in a message-pair.

# H  Examples of Prompts and Messages Written by Ideators

Table 18: Messages generated in our study and the phrase prompt(s) that were shown to ideators.

| Prompt Selection | Phrase(s) Shown | Message Written |
|---|---|---|
| Random(1) | daily club swim workout | Do you want a way to train your whole body? Try a swim workout! You can even join a club to help challenge you to reach your goals! |
| Random(1) | like a barrier of insecurity | Get out and try a new exercise today. Don't let not doing it be a barrier or insecurity. Even pro athletes have to try new exercises for the first time. |
| Directed(1) | snooze button repeatedly isn't exercise | Reminder that hitting the snooze button repeatedly is NOT considered an exercise! Make sure to wake up first thing, and get your legs moving! |
| Directed(1) | next set of stats | Not happy with what you see on the scale or the number of calories you burned? Don't let one day's data ruin your mood. Give it time and you'll see better results if you keep at it! |
| Random(3) | (1) hard workout may feel<br>(2) multiple exercise interventions in terms<br>(3) exercise program for clients plagued | Hard workouts may feel uncomfortable. However, those carry the most enjoyment and success for you! |
| Random(3) | (1) religious institution offers exercise classes<br>(2) workout program because people<br>(3) other forms of water aerobics | Your religious institution offers exercise classes and your local pool offers water aerobics. Exercise with people for motivation! |
| Directed(3) | (1) in the risk of diabetes<br>(2) for the development of diabetes<br>(3) from complications of diabetes | Exercising will help you stay in shape. It will prevent health issues in the future and it can stop the risk of developing diabetes. |
| Directed(3) | (1) book and workout videos<br>(2) mechanics and workout plans<br>(3) exercise tapes or videos | Watching tapes and videos are good ways to try out new exercises. Follow along and impress your loved ones with your new moves! |

# I Thematic Analysis of Messages

Table 19: Themes and categories identified with the qualitative coding of ideated messages.

| Theme | Categories |
| --- | --- |
| Ambiguous Benefits | Ambiguous benefits |
| Anecdote | Anecdote |
| Appeal to "Obvious" Knowledge | Appeal to "obvious" knowledge |
| Appeal by Cohorts | Appeal to children \| Appeal to older ages \| Appeal to overweight |
| Appeal to Fear | Appeal to fear |
| Appeal to Guilt | Appeal to guilt |
| Appeal to Shame | Appeal to shame |
| Appeal to Social Approval | Appeal to social approval |
| Barrier to Ability | Barrier to ability |
| Barrier to Boredom | Encourage exercise variety \| Prompt to try something new \| Tips to make exercise less boring/more fun |
| Barrier to Comfort | Barrier to comfort |
| Barrier to Cost | Cheap exercises \| Lower healthcare/insurance costs |
| Barrier to Effort | It will get easier \| Recommending less effortful routines or exercises \| Take a short break then carry on |
| Barrier to Energy | Barrier to energy |
| Barrier to Enjoyment | Prompt to research fun exercise \| Recommending enjoyable activity |
| Barrier to Motivation | Barrier to motivation |
| Barrier to Resources | At home exercises \| No equipment needed \| No gym available |
| Barrier to Self-Efficacy | Don't feel bad if confused \| Don't need certificate/qualifications \| Improving self-confidence \| Recommending exercises within ability |
| Barrier to Time | Barrier to time |
| Call to Action | Call to action |
| Call to Authority | Citing health experts \| Unspecified authority |
| Collective Societal Benefits | Collective societal benefits |
| Equipment | Bench press \| Exercise machine \| Exercise machines (unspecified \| Rubber exercise tubing/band \| Swimming gear \| Treadmill \| Vertical or horizontal press \| Work - standing desk |
| Exercise Suggestion | Aerobic exercises \| Aerobics \| Anaerobic exercise \| Biking \| Body weight exercises \| Cardiovascular exercise \| Climbing \| Competitive cycling \| Dance \| Diving \| Double clean \| Exercise through chores \| Handstand \| High intensity exercise \| Hot yoga \| Internal rotation workouts \| Iron Yoga \| Jump on bed \| Jumping jacks \| Lift weights \| Lifting luggage \| Meditation \| Pull-ups \| Pushing kids on swings \| Push-ups \| Resistance exercises \| Ring Pull-ups \| Running \| Seated leg-raises \| Sit-ups \| Snatches \| Sports \| Squats \| Strength training \| Strenuous/moderate/vigorous exercise \| Stretching \| Swimming \| Tennis \| Using stairs \| Vertical and horizontal presses \| Volleyball \| Walk your dog \| Walking \| Water exercises \| Work/desk exercise \| Yoga |
| Fear of Injury | Don't overexert yourself \| Recommending exercises to avoid injury \| Research good techniques to avoid injury \| Take breaks \| Tips to avoid injury for outdoor activities |
| Food and Drink | Avoid steroids/pills/drugs \| Avoid unhealthy food \| Exercise supplements \| Exercise to avoid medication/drugs \| Food recommendation \| Staying hydrated \| Stress eating advice |
| Future Life | Improve quality of life \| Live longer |
| Goals | Journaling to track your goal \| Set Actionable goals \| Set daily exercise goal \| Set goals based on health recommendations \| Set unspecified goal \| Set weight goal \| Tips to reach goals \| Visualizing meeting goals |
| Health Advice | Advice for diabetics \| See a doctor if you are worried |

| Theme | Categories |
| --- | --- |
| Health Benefits | Better mobility | Bone health | Cardio Health | Fluid regulation | Help with foot problems | Lowers blood pressure | More stamina | Pain/strain relief | Slows aging process | Strong immune system | Unspecified health benefit |
| Health Risks | Arthritis | Breathing difficulty | Cancer | Depression | Diabetes | Heart disease | Hernias | Obesity | Unspecified health risk |
| Improving Appearance | "Look better" | Beach ready | Chest and back | Improved posture | Look more appealing to potential partners | Nice butt | Six-pack abs |
| Inspirational Phrase | Inspirational phrase |
| Lack of Knowledge | Research exercise routines | Research nutrition | Study exercise form |
| Lack of Social Support | Exercise with an expert | Exercise with friends | Family want you to be healthy | Find places to support you | Impress your doctor | Interacting with others | Join a health club | Join an exercise class | Meeting new people | Playing/exercising with children | Social competition |
| Mental Health Benefits | "Feeling" better | Happier | Improved sleep | Improves cognitive abilities | Lower depression | Lowers anxiety | Relaxing | Release endorphins | Self-esteem | Stress reduction | Unspecified mental health benefits |
| Muscle Building | Biceps | Core strength | Leg Muscles | Physically stronger | Shoulder muscles | Triceps | Unspecified Muscle building | Upper body strength |
| Overcoming Beliefs | Changing mindset |
| Overcoming Family Obligations | Overcoming family obligations |
| Overcoming Self-Consciousness | Overcoming self-consciousness |
| Push to Do More | Prompt to increase | Prompt to steadily increase |
| Push to Start | Prompt to start exercising | Prompt to stop sedentary lifestyle |
| Rewards | Prizes from exercise competitions | Reward with food | Reward with new clothes | Reward with new exercise equipment | Unspecified reward |
| Self-Empowerment | Self-empowerment |
| Self-Forgiving | Self-forgiving |
| Self-Reflection | Self-reflection |
| Social Comparison | Avoid social comparisons | Downwards Social comparison | Upwards Social Comparison |
| Specific Locations | Around the neighbourhood | At desk exercises | At home | At school | Beach | Church/community centre | Front yard | Gym | Outdoors | Park | Travelling/airport | Walk to train station |
| Specificity | Appropriate Exercises (e.g. "Try what's best for you") | Developing habits | Even small amounts of exercise | Exercise daily | Exercise regularly | Follow exercise plan/routine | Specific amount/distance to exercise | Specific days a week to exercise | Specific minutes to exercise |
| Time to See Results | Dedicate time | Fast results | Promise of results | Tips to progress faster |
| Time to Exercise | Anytime | End of the day | Morning exercise | Spring | Summer/hot weather |
| Use of Technology | Exergame | Experts review your exercises from an app | Follow videos | Listen to music/podcast | Reflect on progress | Use apps for exercise tips | Use apps for workout schedules | Use apps to track progress | Watch TV while exercising |
| Weight Loss | Aid digestion | Boost metabolism | Burning calories | Burning cellulite | Maintaining weight | Slimming down |

# J Linear Mixed Models and statistical analysis results of Prompt Creativity, Prompt-Ideation Mediation, and Ideation Diversity

Table 20: Statistical analysis of responses due to effects (one per row), as linear mixed effects models, all with Participant as random effect, Prompt Selection and Prompt Size as fixed effects, their interaction effect. a) model for manipulation check analysis of how prompt configurations affect perceived prompt creativity (RQ1.2); b) model for mediation analysis of how prompt configurations affect ideation effort (RQ2.2). *n.s.* means not significant at p>.01. p>F is the significance level of the fixed effect ANOVA. $R^2$ is the model's coefficient of determination to indicate goodness of fit.

a) Prompt Creativity Manipulation Check (RQ1.2)

| Response | Linear Effects Model (Participant random effect) | p>F | $R^2$ |
|---|---|---|---|
| Prompt Unexpectedness | Prompt Selection + | <.0001 | .523 |
| | Prompt Size + | n.s | |
| | Prompt Selection × Size | n.s | |
| Prompt Understandability | Prompt Selection + | .0008 | .500 |
| | Prompt Size + | <.0001 | |
| | Prompt Selection × Size | .0316 | |
| Prompt Relevance | Prompt Selection + | <.0001 | .450 |
| | Prompt Size + | <.0001 | |
| | Selection × Size | n.s. | |
| Prompt Quality | Prompt Selection + | <.0001 | .572 |
| | Prompt Size + | <.0001 | |
| | Prompt Selection × Size | n.s. | |

b) Prompt-Ideation Effort Mediation Analysis (RQ2.2)

| Response | Linear Effects Model (Participant as random effect) | p>F | $R^2$ |
|---|---|---|---|
| Ideation Fluency | Prompt Selection + | <.0001 | .542 |
| | Prompt Size + | <.0001 | |
| | Prompt Selection × Size | .0042 | |
| Ideation Ease | Prompt Selection + | <.0001 | .546 |
| | Prompt Size + | n.s | |
| | Selection × Size | n.s | |
| Prompt Adoption | Prompt Selection + | <.0001 | .575 |
| | Prompt Size + | n.s | |
| | Prompt Selection × Size | n.s | |

Table 21: Statistical analysis and results of mediation effects (RQ2.3) of how prompt configurations (a) and perceived prompt creativity (b) affect ideation diversity. See Table 20 caption to interpret tables. Positive and negative numbers in second column represent estimated model coefficients indicating how much each fixed effect influences the response.

a) Prompt Distance to Ideation Mediation

| Response | Linear Mixed Effects Model (Participant as random effect) | | p>F | $R^2$ |
|---|---|---|---|---|
| Ideation Mean Pairwise Distance | +0.18 | Prompt Mean Distance + | <.0001 | .399 |
| | +0.06 | Prompt Min Distance + | .0205 | |
| | +0.01 | Pr. P. Chamfer Dist. + | n.s. | |
| | +0.02 | Intra-Pr. P. Mean Dist. | <.0001 | |
| Ideation Minimum Pairwise Distance | +0.10 | Prompt Mean Distance + | .0241 | .315 |
| | +0.15 | Prompt Min Distance + | <.0001 | |
| | +0.06 | Pr. P. Chamfer Dist. + | .0115 | |
| | +0.02 | Intra-Pr. P. Mean Dist. | .0041 | |

b) Prompt Creativity to Ideation Mediation

| Response | Linear Mixed Effects Model (Participant as random effect) | | p>F | $R^2$ |
|---|---|---|---|---|
| Ideation Mean Pairwise Distance | −0.0014 | Pr. Unexpectedness + | .0315 | .367 |
| | +0.0001 | Pr. Understandability + | n.s | |
| | −0.0034 | Prompt Relevance + | .0020 | |
| | +0.0018 | Prompt Quality | n.s | |
| Ideation Minimum Pairwise Distance | +0.0024 | Pr. Unexpectedness + | .0087 | .272 |
| | −0.0026 | Pr. Understandability + | n.s | |
| | −0.0056 | Prompt Relevance + | .0003 | |
| | +0.0052 | Prompt Quality | .0431 | |

**Table 22: Statistical analysis of how prompt selection influences ideation diversity defined by different metrics (RQ3): a) individual diversity, b) collective diversity, and c) thematic diversity. See Table 20 caption for how to interpret tables.**

a) Ideation Individual Diversity

| Response | Linear Effects Model (Participant as random effect) | p>F | $R^2$ |
|---|---|---|---|
| Ideation Mean Pairwise distance | Prompt Selection + | <.0001 | .361 |
| | Prompt Size + | n.s | |
| | Selection × Size | .0005 | |
| Ideation Min Pairwise distance | Prompt Selection + | <.0001 | .296 |
| | Prompt Size + | n.s | |
| | Selection × Size | n.s | |
| Ideation Self-Quality | Prompt Selection + | n.s | .570 |
| | Prompt Size + | .0292 | |
| | Selection × Size | .0152 | |

b) Ideation Collective Diversity

| Response | Linear Effects Model (Sample as random effect) | p>F | $R^2$ |
|---|---|---|---|
| Ideation Dispersion | Prompt Selection + | <.0001 | .873 |
| | Prompt Size + | n.s | |
| | Prompt Selection × Size | <.0001 | |
| Ideation Evenness | Prompt Selection + | <.0001 | .984 |
| | Prompt Size + | .0030 | |
| | Prompt Selection × Size | .0061 | |

c) Ideation Collective Diversity (Thematic Coding)

| Response | Linear Effects Model (Sample as random effect) | p>F | $R^2$ |
|---|---|---|---|
| Category Flexibility | Prompt Selection | <.0001 | .979 |
| Category Originality | Prompt Selection | <.0001 | .933 |
| Theme Flexibility | Prompt Selection | <.0001 | .911 |
| Theme Originality | Prompt Selection | <.0001 | .396 |

**Table 23: Statistical analysis of how prompt selection influences ideation creativity as validated by different methods (RQ3.1): a) individual rating, b) collective ranking, and c) collective pairwise rating. See Table 20 caption for how to interpret tables.**

a) Individual Rating Validation

| Response | Linear Effects Model (Participant + Ideation as random effects) | p>F | $R^2$ |
|---|---|---|---|
| Ideation Informative Helpfulness | Prompt Selection | <.0001 | .559 |
| Ideation Quality | Prompt Selection | n.s. | .467 |

b) Collective Ranking Validation

| Response | Linear Effects Model | p>F | $R^2$ |
|---|---|---|---|
| Ideations Unrepetitive | Prompt Selection | <.0001 | .284 |
| Ideations Informative | Prompt Selection | <.0001 | .340 |
| Ideations Motivating | Prompt Selection | .0426 | .028 |

c) Collective Pairwise Rating Validation

| Response | Linear Effects Model | p>F | $R^2$ |
|---|---|---|---|
| Difference Rating | Prompt Selection | <.0001 | .279 |

## K  Investigating Confound of Prompt Understandability on Ideation Diversity

Having found that prompt understanding difficulty is correlated with ideation diversity, we investigated the alternative hypothesis that the difficulty of interpreting the prompts was a key reason for improved ideation because of increased ideation determination, rather than the content diversity in phrases due to the prompt selection technique. We argue that the increase in ideation diversity due to Directed Diversity is evidenced by increased perceived diversity ratings from validators and the higher number of idea categories from the thematic analysis. This shows that Directed Diversity did stimulate more diverse ideas due to some knowledge transfer from prompt to ideations, albeit with difficulty. We identify three more sources of evidence next.

First, we qualitatively analyzed ideation rationales and found that while prompts could be rated hard to understand or irrelevant, participants still adopted some ideas. Ideators cherry-picked parts that were usable or conceived tangential ideas: e.g., P1 read *"orthopaedic surgeons and exercise specialists"* and decided to *"cut out the bit about surgeons... I focused on the idea of specialists..."*; P2 read *"ballistic stretch uses vigorous momentum"*, commented that *"this isn't a phrase that I'm familiar with"*, yet could write about stretching: *"Stretch, breathe, and feel mindful."*

Second, we quantitatively analyzed the Ideation Mean Pairwise Distance for prompts that participants understood (Phrase Understanding factor > 0). Table 24a describes the statistical analysis of the linear mixed effects model. We found that although distance was slightly higher (i.e., less diverse) when ideators understood phrases less, regardless of understanding, ideations from Directed(1) prompts had higher distances than ideations from Random(1) prompts (Figure 25, left). The effect due to Prompt Type was larger than due to Phrase Understanding. Furthermore, we analyzed whether the difficulty to understand may manifest as slower ideation speed due to more thinking time to ideate to lead to better diversity, but did not find a correlation between phrase understanding and ideation speed ($\rho = .046, p = n.s.$), and found the opposite effect that slower ideations led to lower distances (Table 24a and Figure 25 right). These suggest that prompt selection is a primary factor.

Third, we investigated if Directed Diversity helped to stimulate ideas closer to prompts than would be done naturally without prompts (None) or accidentally with Random prompts. We analyzed this by calculating the prompt-ideation distance between Directed prompts and their corresponding ideated message, and their closest None and Random messages. Table 24b describes the statistical analysis of the linear mixed effects model. Figure 26 shows that the directed ideations were closest to the prompts, indicating the efficacy of Directed Diversity to transfer knowledge for ideation diversity.

Table 24: Statistical analysis of a) how ideators' understanding of phrases influences ideation diversity and b) how similar Directed ideations are to their prompts compared to other None and Random Messages.

| | a) Ideation Individual Diversity | | | | b) Prompt-Ideation Closeness | | |
|---|---|---|---|---|---|---|---|
| Response | Linear Effects Model (Participant as random effect) | p>F | $R^2$ | Response | Linear Effects Model (Participant random effect) | p>F | $R^2$ |
| Ideation Mean Pairwise distance | Prompt Selection + | <.0001 | .289 | Prompt-Message Distance | Message Type | <.0001 | .047 |
| | Prompt Size + | n.s | | | | | |
| | Selection × Size + | .0001 | | | | | |
| | Phrase Understanding>0 + | .0074 | | | | | |
| | Selection × (Understanding>0) + | n.s. | | | | | |
| | Log(Ideation Speed)>Median + | .0043 | | | | | |
| | Selection × (Speed>Median) | .0213 | | | | | |

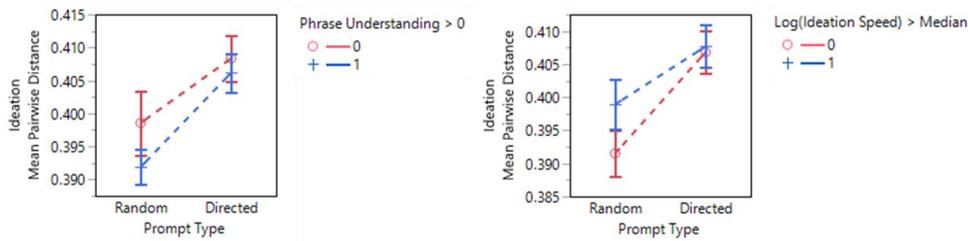

Figure 25: Results of computed individual diversity from ideations for different prompt configurations for (left) prompts that users understood (>0) or did not and (right) ideations that were fast or slow.

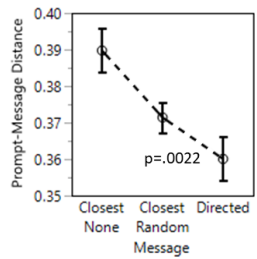

Figure 26: Results of prompt-message distance (how dissimilar a prompt is from a message) comparing different messages with respect to Directed(3) prompts.

## L  Examples of Message-Group Ranking

The factors of message-group ranking were derived from the sum of rankings (for each of the three condition) per validator for his five ranking trials (see the factor loadings in Table 15: The rotated factor loading of factor analysis on metrics of group ranking of the generated messages.). Therefore, these factors reflect the probability of how a validator ranked the message-groups of each condition. The following table shows examples of the factors and the corresponding message-group samples.

**Table 25: Examples of the factors with low (< Median) and high (≥ Median) scores for "Ideations Unrepetitive" and "Ideations Informative".**

| Ideations Unrepetitive | Ideations Informative | Example Message-Group of 5 Ideations |
|---|---|---|
| High | High | • Exercise can help you have really good sleep.<br>• Why don't you try something new? Shake it up a little? Maybe lift a few small weights, or add in some squats - variety keeps things interesting.<br>• You have 24 hours in a day-- think about how much time you spend on social media or doing something that's not going to benefit you in the long run and use that time to workout by prioritizing your health!<br>• Go for the goal, do not stop, do not think you cannot do it. YOU CAN!<br>• Summer is coming up and you want to look good when you are outside. Exercising at a health club is a good way to meet other people. Have a friend to work out with you and have each other motivate each other. |
| High | Low | • Just get moving. It's that simple.<br>• Your dog is bored. Take him for a walk! It's good for both of you and he'll be thrilled!<br>• Work out more. You will feel and look better. You will get more toned.<br>• Exercising can improve your cardio health, thus helping you to live a more fulfilling life.<br>• Start exercising more! You'll improve your mood and boost your self confidence. You'll feel great! |
| Low | High | • Switch off an air conditioner while working out. Let the sweat out, and burn some calories.<br>• Not happy with what you see on the scale or the number of calories you burned? Don't let one day's data ruin your mood. Give it time and you'll see better results if you keep at it!<br>• Sleep is when the body recovers and is very important. Rest early and run tomorrow!<br>• Overcome your anger and your fear by going to the gym and working out!<br>• Always stretch so that you perform at your best. You can do it! |
| Low | Low | • Exercise helps build strong muscles as well as well as making your body more flexible. You will reduce your risk of disease and injury by keeping up with your program.<br>• The first page of every book is the hardest to grasp, the first drink tastes the most sour and the first minute of every exercise is the hardest. All things get easier as you press on.<br>• Walking to the train station is better as it gets you more active. Avoid lifts to the train station.<br>• Keep exercising to keep your mind off difficult personal issues, like college admissions.<br>• By using the proper squat position, you can train muscles that take pressure off of your knee and back to help with pain in both areas. |